\documentclass[12pt]{iopart}
\usepackage{graphicx}
\usepackage{iopams} 
\usepackage[colorlinks,bookmarks=false,citecolor=blue,linkcolor=red,urlcolor=blue]{hyperref}
\usepackage{epstopdf}

\usepackage{color} %,amsmath}
\newcommand{\bc}{\begin{center}}
\newcommand{\ec}{\end{center}}
\newcommand{\be}{\begin{equation}}
\newcommand{\ee}{\end{equation}}
\newcommand{\bea}{\begin{eqnarray}}
\newcommand{\eea}{\end{eqnarray}}

\definecolor{todocolor}{rgb}{0.8,0.1,0.2}

\newcommand{\bss}[1]{\mathbf{#1}}
\newcommand{\comm}[2]{\left[#1,#2\right]}

\newcommand{\ket}[1]{\left|#1\right\rangle}
\newcommand{\bra}[1]{\left\langle#1\right|}

\newcommand{\up}{\uparrow}

\newcommand{\eps}{\varepsilon}

\newcommand{\eqref}[1]{(\ref{#1})}

\newcommand{\BS}{\bss{S}}
\newcommand\nd{^{\vphantom{\dagger}}}
\newcommand\yd{^\dagger}

\def\ie{\emph{i.e.},\ }
\def\eg{\emph{e.g.}\ }

\begin{document}

%\title{Properties of entanglement spectra in quantum $S = 1$ chains}
%\title{Quantum $S=1$ chain: an unorthodox entanglement perspective}
\title{Entanglement analysis of isotropic spin-1 chains}

\author{Ronny Thomale$^1$, Stephan Rachel$^2$, B. Andrei Bernevig$^3$, and Daniel P. Arovas$^4$}

\date{\today}

\address{$^1$ Institute for Theoretical Physics I, University of W\"urzburg, D-97074 W\"urzburg, Germany} 
\address{$^2$ Institute for Theoretical Physics, TU Dresden, 01062 Dresden, Germany} 
\address{$^3$ Department of Physics, Princeton University, Princeton, NJ 08544, USA}
\address{$^4$ Department of Physics, University of California at San Diego, La Jolla, California 92093, USA} 

\begin{abstract}
We investigate entanglement spectra of the SO(3) bilinear-biquadratic spin-1 chain, a model with phases exhibiting spontaneous
symmetry breaking (both translation and spin rotation), points of enlarged symmetry, and a symmetry-protected topological phase (the Haldane phase). Our analysis reveals how these hallmark features are manifested in the entanglement spectra, and highlights the versatility of entanglement spectra as a tool to study one-dimensional quantum systems via small finite size
realizations.
\end{abstract}

%\pacs{03.67.Mn, 05.30.-d, 05.70.Jk, 71.10.Pm}
%\noindent{\it Keywords}: Entanglement entropy
%\submitto{J. Stat. Mech.}

%% \maketitle
%\tableofcontents

\section{Introduction}
The notion of entanglement is central to quantum information theory, and has been applied with great success in the analysis
of condensed matter systems \cite{RevModPhys.80.517}.  Von Neumann entanglement entropy and its generalizations have been
invoked in various contexts, including thermalization, diagnosis of topological phases, and properties of conformally invariant
quantum systems.  Beyond entanglement entropy, the full `entanglement spectrum' of a system's reduced density matrix $\rho\nd_A$
reveals much detailed information about the structure of its ground state \cite{li-08prl010504}.  In particular, block-diagonal structures in $\rho\nd_A$ derived
from a many-body ground state projector $\ket{\Psi_0}\bra{\Psi_0}$ may be exploited
to analyze entanglement spectra as a function of quantum numbers associated with symmetries which are preserved by the entanglement cut. 
The notion of an `entanglement gap' was established~\cite{li-08prl010504,thomale-10prl180502} to suggest a separation of  ``relevant'' low-lying entanglement levels from
``spurious'' low probability (high entanglement energy) levels.
In some cases, such as the fractional quantum Hall effect and symmetry-protected topological phases (SPTs)~\cite{thomale-10prl180502,PhysRevLett.107.157001,PhysRevLett.106.100405,PhysRevB.84.205136,xlqi,PhysRevB.82.241102,PhysRevB.84.045127,PhysRevB.86.041401,franky,PhysRevLett.104.130502}, the relevant low-lying levels have similarities to the energy levels of an edge Hamiltonian, although this line of reasoning is not universal \cite{chandran-arxiv:1311}.

While the task of connecting entanglement measures to physical observables is still at an early stage~\cite{PhysRevB.85.035409}, the entanglement spectrum in particular has become a major tool to investigate one-dimensional quantum many-body systems. On a qualitative level, this is due to the fact that entanglement spectra allow the reliable extraction of information far beyond that obtained from the entanglement entropy. For gapless 1-d systems, the entanglement spectral distribution is fully characterized by the central charge of the associated conformal field theory~\cite{caprese} which one could deduce from the entanglement entropy~\cite{calacardy}. The entanglement spectrum, however, also determines the boson compactification radius associated with the operator content of the boundary CFT~\cite{lauch}. For gapped 1-d systems the entanglement spectrum can completely determine whether a system is topological or not, and provides a classification of such topological 1-d phases \cite{franky,PhysRevB.83.075103}. 

In this article we investigate the finite size structure of the entanglement spectrum of the spin-1 bilinear biquadratic model for the spatial ground state bipartition of a 12 site realization, which will be defined in Section~\ref{sec:model}. For the ferromagnetic phase exhibiting spontaneous spin rotation symmetry breaking in the thermodynamic limit, the entanglement spectrum is calculated analytically and interpreted with respect to finite vs.\ infinite size entanglement in Section~\ref{sec:ferro}. In Section~\ref{sec:quadru}, the Uimin-Lai-Sutherland point of the quadrupolar phase is shown to exhibit an extensive number of zeroes in the reduced density matrix which we analytically trace back to its enlarged internal SU(3) symmetry. In Section~\ref{sec:dimer}, the dimerized phase with mod2 translation symmetry breaking is readily identified by the even-odd asymmetry of the ground state bipartition. At the negative biquadratic point, breaking of external symmetries, i.e. symmetry transformations involving the spatial coordinates such as translation, together with internal symmetry enhancement is shown to yield a rich entanglement structure. In the symmetry-protected topological Haldane phase discussed in Section~\ref{sec:haldane}, we find a correspondence between entanglement spectra and the low-energy spectrum of an open boundary Hamiltonian. We further employ the single mode approximation to obtain perturbative ground states around the Affleck-Kennedy-Lieb-Tasaki (AKLT) point, and show that it correctly yields the lowest-lying entanglement deformations away from AKLT. In Section~\ref{sec:conclusion}, we conclude that entanglement spectra for finite size realizations constitute a highly efficient tool to understand the underlying structure of one-dimensional quantum many-body systems.

\section{Model and ground state bipartition}
\label{sec:model}

\begin{figure}[t]
\centering
\includegraphics[width=0.95\columnwidth,clip]{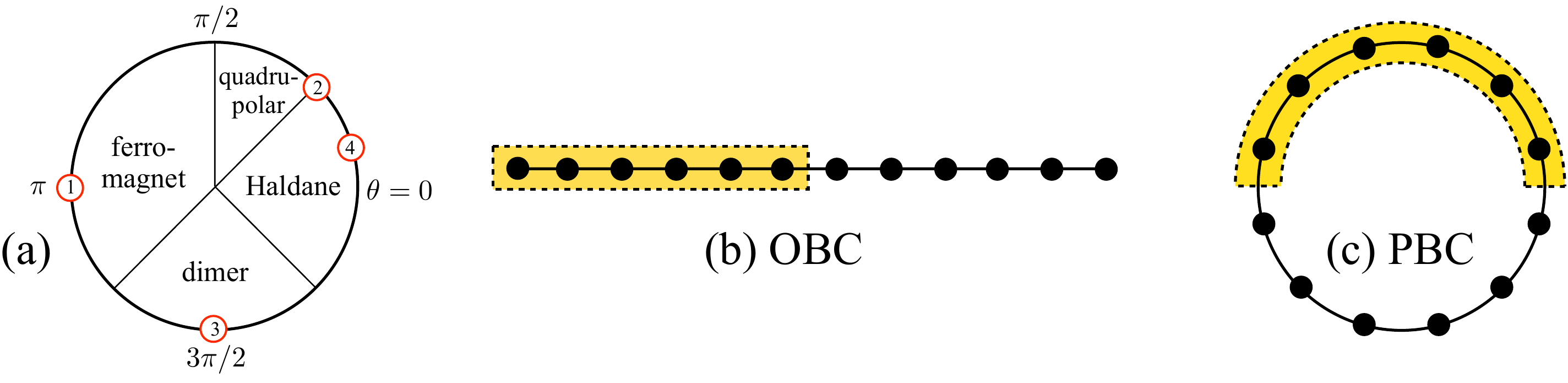}
\caption{(a) Schematic phase diagram of the bilinear biquadratic model~\eqref{ham}. The four points highlighted by red circles are of central interest in the following: (1) Heisenberg ferromagnet, (2) Uimin-Lai-Sutherland point, (3) the negative biquadratic model, and (4) the AKLT point.
(b) Spatial bipartition of an open boundary $12$ site chain resulting in one single cut. (c) Spatial bipartition of a periodic $12$ site chain resulting in two cuts.}
\label{fig:cut}
\end{figure}
%Brief but complete discussion of the phase diagram of the BLBQ spin 1 chain.
We consider the bilinear biquadratic spin 1 chain given by 
\begin{equation}
H=\sum_n \cos\theta\>\BS_n\cdot\BS_{n+1} + \sin \theta\, \big(  \BS_n\cdot\BS_{n+1} \big)^2\ ,
\label{ham} 
\end{equation}
where $\BS_n=(S_n^x, S_n^y, S_n^z)$ are SO(3) spin operators $[S_n^\alpha, S_{n'}^\beta]=i \hbar \epsilon_{\alpha \beta \gamma} S_n^\gamma\,\delta_{nn'}$ and  the angular parameter
$0 \leq \theta < 2\pi$ specifies the sign and relative strength of the bilinear and biquadratic interactions. The phase diagram has been thoroughly investigated by numerical studies (Fig.~\ref{fig:cut}).
A ferromagnetic phase for $\pi/2 < \theta < 5 \pi/4$ exhibits spin rotation symmetry breaking and is the only phase of~\eqref{ham} in which the ground state does not reside in the singlet sector.
The phase for $\pi/4 < \theta < \pi/2$ has been investigated numerically \cite{fath1991} and was found to be gapless with soft modes at $k=0$, $2\pi/3$ and $4\pi/3$,
with the dominant quadrupolar correlations~\cite{itoi,trebst}.   At $\theta=\pi/4$, the quadrupolar phase terminates in the integrable Uimin-Lai-Sutherland point~\cite{uimin,lai,sutherland}, which
will be discussed further below.  Throughout $-\pi/4 < \theta < \pi/4$, the ground state is a symmetry-protected topological phase identified by Haldane~\cite{haldane-phase}.  Anchored by
the exact solution for a gapped valence bond solid at the AKLT point $\tan\theta={1\over 4}$ \cite{aklt}, the persistence of the spin gap throughout the Haldane phase was investigated and
confirmed by early large scale numerics~\cite{white,schollwoeck}. Finally, the dimerized phase for $5 \pi /4 < \theta < 7\pi /4$ exhibits translation symmetry breaking as well as an associated dimerization gap, and is framed by the multicritical SU$(2)_2$ Takhtajan-Babujian point~\cite{takhtajan,babujian,affleck-haldane} to the Haldane phase and by the negative Uimin-Lai-Sutherland
point to the ferromagnetic phase~\cite{buchta}, where the latter transition point is challenging to resolve numerically because of the rapidly vanishing dimerization gap~\cite{hu-arxiv1401}.

Consider a finite size realization of~\eqref{ham} for a given $\theta^*$ for $N$ sites on open (OBC) or periodic (PBC) boundary conditions (Fig.~\ref{fig:cut}). Because of finite size splitting, there will always be a unique ground state $\ket{\Psi_0^{\theta^*}}$ located in a Hilbert space $\mathcal{H}= \bss{1} \times \bss{1} \times \bss{1} ... \times \bss{1}= \otimes_1^N \bss{1}$, where $\bss{1}$ represents the local spin-1 degree of freedom at a single site. According to the Schmidt decomposition with respect to a spatial bipartition, we split the Hilbert space into two regions $A$ and $B$ according to $\mathcal{H}=\mathcal{H}_A \times \mathcal{H}_B$, with the number of sites included in the regions given by $N_A+N_B=N$. From there, we define the reduced density matrix $\rho\nd_A=$Tr$_{B}\rho$, where $\rho=\ket{\Psi_0^{\theta^*}}\bra{\Psi_0^{\theta^*}}$, and Tr$_{B}$ denotes a partial trace over all $N_B$ sites. Following~\cite{li-08prl010504}, this is associated with a Boltzmann factor of unit temperature 
$\rho\nd_A = \exp (-H_A)$, where $H_A$ denotes the entanglement Hamiltonian of region $A$. Depending on OBC or PBC, $H_A$ mimics one or two cuts between region $A$ and $B$ (Fig.~\ref{fig:cut}). The entanglement spectral levels will be denoted by $\xi=\;$spec$[H_A]=-\log($spec$[\rho\nd_A])$. (Note that aside from the real space partition introduced above, other cuts such as the momentum cut~\cite{thomale-mom1,lundgren-arxiv:1404,rex} or the rung cut~\cite{poilblanc,PhysRevB.86.224422} have been advocated in the context of low-dimensional spin systems.) 

 In the case of an isotropic spin chain as in~\eqref{ham}, the block diagonal structure of $\rho\nd_A$ (and equivalently $H_A$) is fully determined by the spin representation of $\ket{\Psi_0^{\theta^*}}$. If it is a singlet, i.e. $S^\alpha \ket{\Psi_0^{\theta^*}}=0$, it yields $[S_A^\alpha, \rho\nd_A]=0$ , where $S_A^\alpha=\sum_{i \in A} S_i^\alpha$, $\alpha=x,y,z$, which gives the SO(3) multiplet structure of $[\BS_A^2,\rho\nd_A]=0$ characterized by the eigenvalues $S_A(S_A+1)$.  If $\ket{\Psi_0^{\theta^*}}$ is a spin multiplet such as in the ferromagnetic phase, it only holds $[S_A^z, \rho\nd_A]=0$ which gives a U(1) symmetry rather than a full SO(3) multiplet structure. In what follows, $N_A$ and $S_A$ characterize the subblocks of $H_A$. These quantum numbers allow to resolve manifestations of symmetry breaking, enhanced internal symmetry, and symmetry-protected character from the sole knowledge of finite size ground states of~\eqref{ham}. 
%%%%The standard scenario investigated will be a $N=12$ chain.

%Definition of entanglement spectrum (ES), reduced density matrix, real space cuts.  Explain how to calculate it explicitly and how to compute it within ED and DMRG. Show that $\rho\nd_A$ commutes with $\vec S_A^2$ and with $S_A^z$ if and only if the Hamiltonian $H$ commutes with $\vec S^2$ and $S^z$. Generalize to SU(3) generators $J^3$ and $J^8$ and the Casimir operator $\vec J^2$.
%Discuss that such considerations imply that the entanglement Hamiltonian in region $A$ might be related (or even identical) to the real Hamiltonian with length $\ell_A$ with OBC.

%\section{ENTANGLEMENT VIEW ON THE SPIN 1 CHAIN}
%As this is supposed to be an unorthodox entanglement perspective we start the discussion with the ferromagnet (instead of the Haldane phase as everyone else does).

%Here we should review the main results and most important papers.
%Maybe we could then discuss  the difference between gapped and critical chains for the entanglement gap scaling. Compare with Calabrese-Lefevre (only very brief - you don't want to go into details).
%\begin{figure}[h!]
%\centering
%\includegraphics[width=0.95\columnwidth,clip]{entgap-scaling}
%\caption{Scaling of the entanglement gaps for both critical and gapped chains for both OBC and PBC.}
%\label{fig:scaling}
%\end{figure}

\section{Ferromagnetic phase: spin rotation symmetry breaking}
\label{sec:ferro}

The ferromagnet extends to a wide region in~\eqref{ham}, hosting at its boundaries at $\theta=\pi/2$ and $\theta=5\pi/4$ two SU(3) symmetric ferromagnets. In many respects, the ferromagnet is the most trivial phase from the viewpoint of entanglement. In the thermodynamic limit were a magnetization axis is spontaneously chosen, the ground state is a single site product state and does not contain any entanglement. For finite size, however, the continuous spin symmetry of~\eqref{ham} cannot be broken, whereas the ground state $\ket{\Psi_{\rm FM}}$ picks the largest available SO(3) representation. As such, $\ket{\Psi_{\rm FM}}$  is uniquely defined independent of the specific choice of Hamiltonian in the ferromagnetic phase. As elaborated on in Section~\ref{sec:model}, $\ket{\Psi_{\rm FM}}$ retains a U(1) symmetry in the entanglement spectrum $\rho\nd_A$ such that we can decompose it into according subblocks. We restrict the following discussion to $S^z\geq 0$ ($S_A^z\geq 0$) as the U(1) symmetry guarantees that subblocks with $-S^z$ ($-S^z_A$) are identical to those with $S^z$ ($S^z_A$).

%The ferromagnetic ground state of a spin-${1\over 2}$ chain takes the simple form of equal occupation of all basis states independent of the $S^z$ sector. For a spin 1 chain considered here, this has to be slightly adjusted. 
%%%%%%%%%%%%%%%%%%%%%%%%%%%%%%%%%%%%%%%%%%%%%%%
\begin{figure}[t]
\centering
\includegraphics[width=0.99\columnwidth,clip]{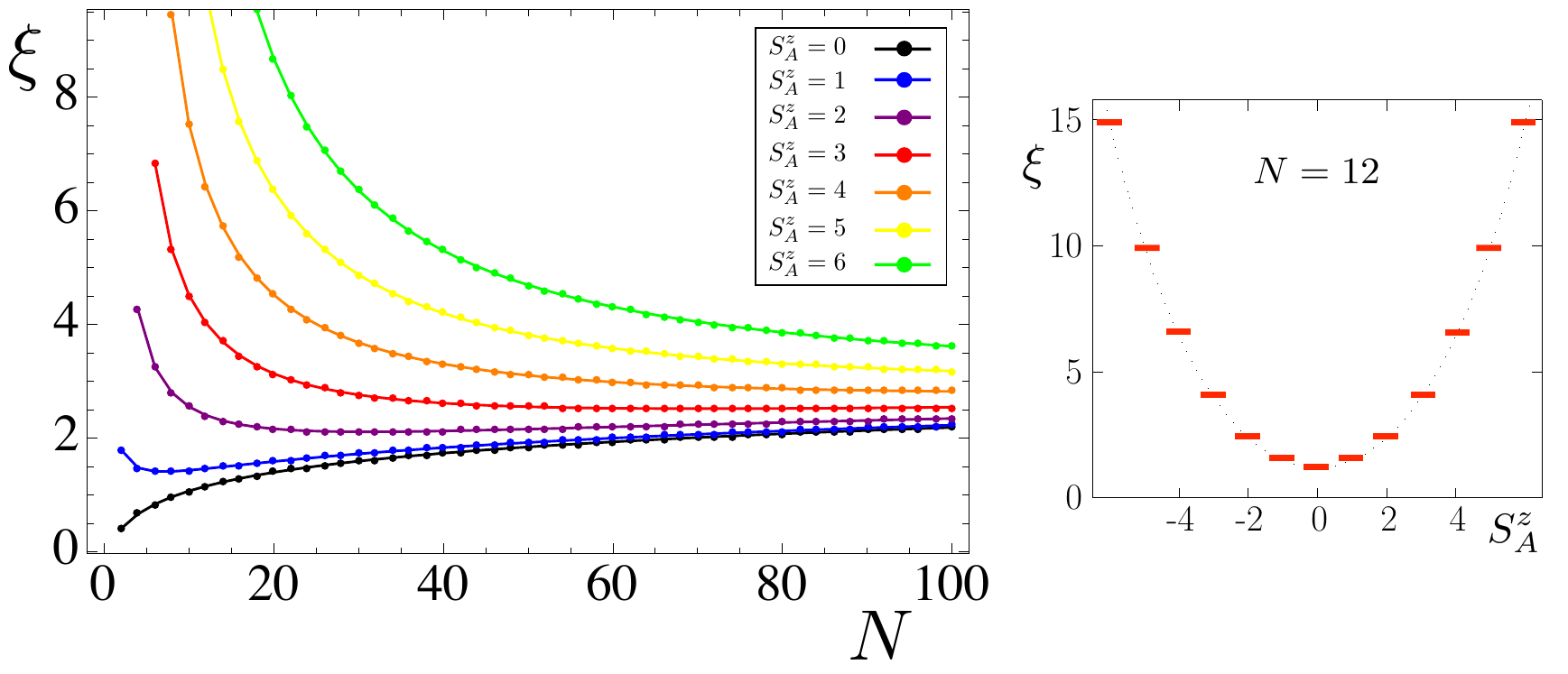}
\caption{(Left) Analytical entanglement spectrum of the spin 1 ferromagnet as a function of total system size $N$ for PBC. Each $S_A^z$ sector contains one finite entanglement level. Different colors correspond to different $S_A^z$ sectors as indicated in the legend. (Right) Example: ES vs.\ $S_A^z$ for $N=12$.}
\label{fig:es-fm}
\end{figure}
%%%%%%%%%%%%%%%%%%%%%%%%%%%%%%%%%%%%%%%%%%%%%%%

We begin with the ferromagnetic spin 1 ground state wave function. In the spin-${1\over 2}$ case, the wavefunction has the most simple form, as all basis states acquire identical weight. For spin 1, we start by the subsector with maximum magnetization $S^z = N$, $\ket{\Psi_{\rm FM}^{S^z=N}} = \ket{\up\up \ldots \up}$. From the knowledge of the multiplet structure, it must hold
$S^2 (S^-)^m \ket{\Psi_{\rm FM}^{S^z=S}} = S(S+1)(S^-)^m \ket{\Psi_{\rm FM}^{S^z=S}} \;\; \forall  \; m \, \in[0,\ldots,2S]$. This allows us to compute the relative weights of the different wave function partitions in $S^z$ basis. With respect to $\ket{\Psi_{\rm FM}^{S^z=N}}$, the spin~1 ferromagnetic wave function configurations acquire different relative weights due to different multiplicities of spin flip operators acting on the same site. For example, given $S^z = N-2$, the basis is constructed by applying $S_i^- S_j^- \ket{\up\up\ldots\up}$, and configurations with $i\not= j$ acquire twice the weight of $i=j$ due to bosonic normalization.
%The reason why the ferromagnetic spin 1 wavefunction does not have the same relative weights (in contrast to the spin-${1\over 2}$ case) for all basis states, can be most conveniently understood by considering the sector with third-highest magnetization, $S^z = L-2$. The basis can be constructed by applying twice the lowering operator on the fully polarized state, $S_i^- S_j^- \ket{\up\up\ldots\up}$. Now we have to distinguish between the cases where $i\not= j$ and $i=j$ as the relative weight in the latter case is half the weight of the former case. 
%For smaller $S^z$ subspaces there are more possibilities to distribute the lowering operators leading to more different relative weights.

Consider the bipartition of $\ket{\Psi_{\rm FM}}$ in the $S^z=0$ sector with $N_A=N/2$, as illustrated in Fig.~\ref{fig:cut}.  In the following, we assume the Schmidt basis of region $A$ and $B$ to be ordered according to decreasing relative weight. From the Schmidt decomposition, the (in general rectangular) entanglement matrix $E_A$ can be defined as $\rho\nd_A = E_A^T E_A$, which for $S_A^z=0$, aside from global normalization, takes the form  
\begin{equation}
E_A = \left( \begin{array}{cccccccc}
1 & 1/2 & 1/2 & \cdots & 1/4 & 1/4 & \cdots & (1/2)^m \\[5pt]
1/2 & 1/4 & 1/4 & \cdots & 1/8 & 1/8 & \cdots & (1/2)^{m+1} \\[5pt]
1/2 & 1/4 & 1/4 & \cdots & 1/8 & 1/8 & \cdots & (1/2)^{m+1} \\
\vdots &&&&&&& \vdots
\end{array}\right)\ .
\end{equation}
One finds that the matrix $E_A$ consists of several blocks with identical individual entries, the size of which being dictated by the number of basis states with the same relative weight. Determining the size of these blocks is given by the dimension of the $S^z$ ($S_A^z$) subspaces of the  Hilbert space for a given chain length $N$ ($N_A$),
\begin{equation}
{\rm dim}\,\mathcal{H}( N, S^z ) = \hskip-0.5cm\sum_{j=0}^{\lfloor (N-S^z)/2 \rfloor} \left( N \atop {S^z + 2j} \right)\left( S^z + 2j \atop j \right)\ ,
\end{equation}
where $\lfloor x \rfloor$ denotes the floor function. 
%Note that the sum over the two binomials can be expressed as the hypergeometric function ${}_2F_1$. Now let us analyze the analytical structure of the matrix $E_A$. 
By construction, all columns (and rows) are linearly dependent, \ie $E_A$ has rank 1 and the characteristic polynomial reads $P[\mu] = \mu^{N_A-1} (\mu - g)$ where $\mu=g$ is the only non-zero eigenvalue given by
\begin{equation}\label{formula-fm-es}
g\nd_{[N_A, S_A^z]} =\hskip-0.5cm \sum_{j=0}^{\lfloor (N_A-S^z_A)/2 \rfloor}\left( N_A \atop {S_A^z + 2j} \right)\left( S_A^z + 2j \atop j \right) 2^{-(2j+S_A^z)}.
\end{equation}
Restoring the overall normalization of $\ket{\Psi_{\rm FM}}$ leads to
$\tilde g\nd_{[N_A, S_A^z]} = g\nd_{[N_A, S_A^z]}\big/\sqrt{\mathcal{N}_{[N, S^z_{\rm tot}]}}\, $, where the norm is given by
\begin{equation}
\mathcal{N}_{[N, S^z_{\rm tot}]}^2 = \hskip-0.5cm\sum_{j=0}^{\lfloor (N-S^z_{\rm tot})/2 \rfloor} \left( N \atop {S^z_{\rm tot} + 2j} \right)\left( S^z_{\rm tot} + 2j \atop j \right)
 2^{-2j}\ .
\end{equation}
The non-zero eigenvalue of the reduced density matrix is then given by $\left| \tilde g\nd_{[N_A, S_A^z]}  \right|^2$, where the entanglement spectrum $\xi=-\log \big| \tilde g\nd_{[N_A, S_A^z]}  \big|^2$ is visualized for large system sizes in Fig.\,\ref{fig:es-fm}. In the right panel of Fig.\,\ref{fig:es-fm}, the same ES for $N=12$ is plotted as a function of $S_A^z$. The parabolic form is a direct consequence of the binomial Hilbert space dimension for the different $S_A^z$ sectors.

\section{Quadrupolar phase: enhanced SU(3) symmetry at $\theta=\pi/4$}
\label{sec:quadru}

Following numerical work \cite{fath1991}, it was understood that the quadrupolar phase could be suitably characterized by the $\theta=\pi/4$ point. This is the Uimin-Lai-Sutherland (ULS) model,
an integrable SU$(3)_1$ model with an exact solution of the Bethe {\it Ansatz\/} type \cite{uimin,lai,sutherland}.   At this point, the interaction between sites is, up to a constant, proportional
to $-{\rm P}\nd_1(n,n+1)$, where ${\rm P}\nd_1$ is the projector onto total spin $J=1$ for the neighboring sites.  The $J=1$ triplet is thus the ground state of the link, and the
$J=0$ singlet and $J=2$ quintuplet combine to form a degenerate sextuplet.  The Hamiltonian
can then be mapped onto that of the antiferromagnetic SU(3) Heisenberg chain in the fundamental representation, since $\bss{3}\otimes\bss{3}=\bss{\bar 3}\oplus\bss{6}$.
As known from the SU($N$) formulation of the Lieb-Schultz Mattis theorem~\cite{lsm}, and consistent with the generalized SU($N$) Haldane-gap criterion~\cite{confine}, this model is gapless. Despite its solvability, from the viewpoint of numerical entanglement measures, it is still generally involved to extract the critical theory content for such models from finite
size scaling~\cite{max}.

\begin{figure}[t]
\centering
\includegraphics[width=0.50\columnwidth,clip]{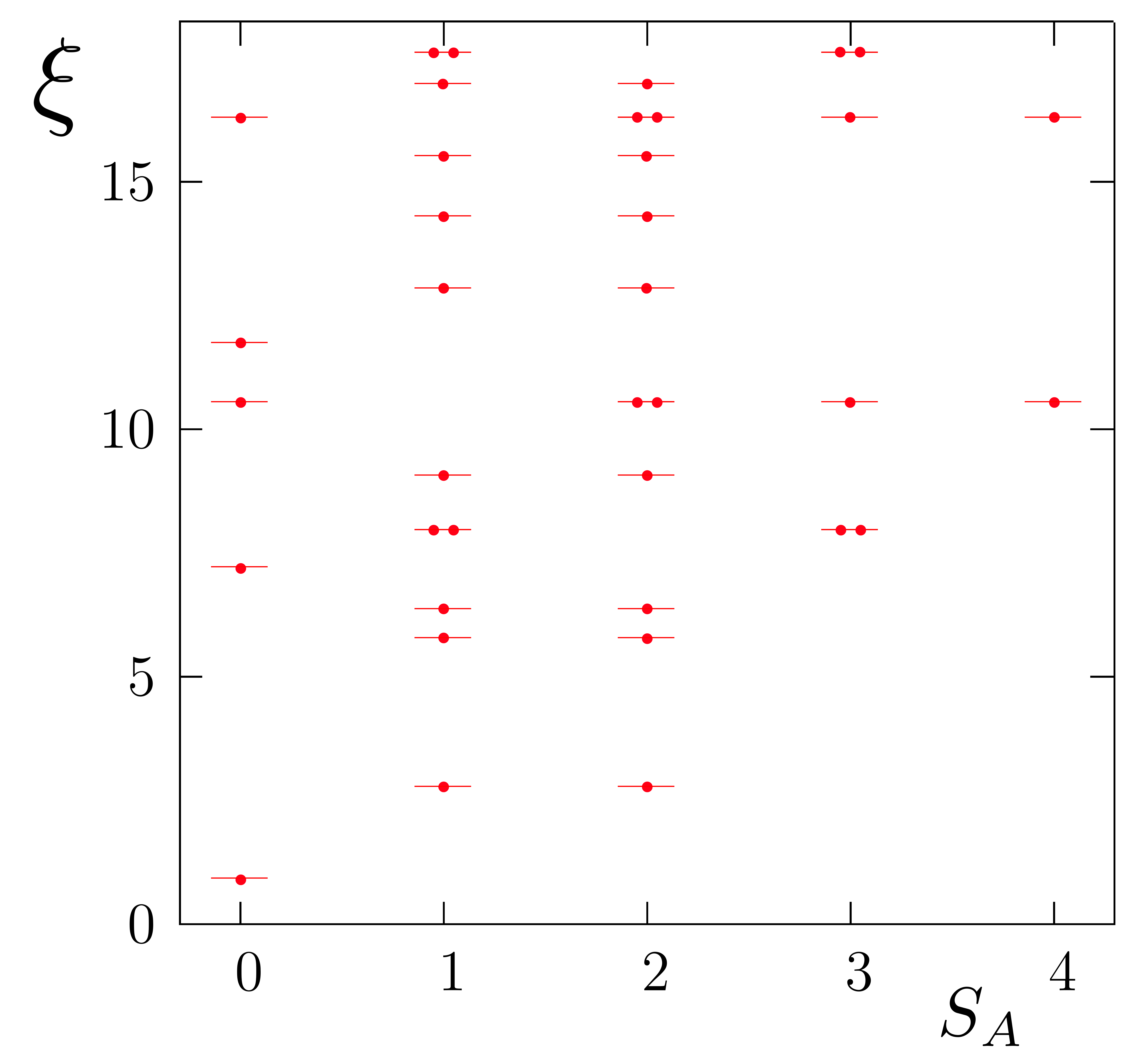}
\caption{$(N,N_A)=(12,6)$ entanglement spectrum (PBC) of the Uimin-Lai-Sutherland point decomposed into SO(3) representations labeled by $S_A$. The enhanced SU(3) symmetry becomes apparent from the eigenvalue degeneracies of different SO(3) representations, such as a triplet and quintuplet forming an SU(3) octet as the first entanglement energy above the lowest singlet level.}
\label{fig:ULS}
\end{figure}

The enhanced internal SU(3) symmetry, however, can be readily derived (see also~\ref{app:lambda}) and, because of its locality, manifests itself for any finite size realization. Along with the SU(3) singlet property of the ground state, this yields an SU(3) multiplet structure for $\rho\nd_A$, where the representations of SU(2) arrange accordingly in the entanglement spectrum (Fig.~\ref{fig:ULS}). (Alternatively, one can directly adapt SU(3) Casimir operators $J_3$ and $J_8$ to label the individual blocks of $\rho\nd_A$.) Aside from the exact SU(3) symmetry specific to the ULS point, the qualitative entanglement features carry over to the whole quadrupolar phase.

\begin{figure}[t]
\centering
\includegraphics[scale=0.5]{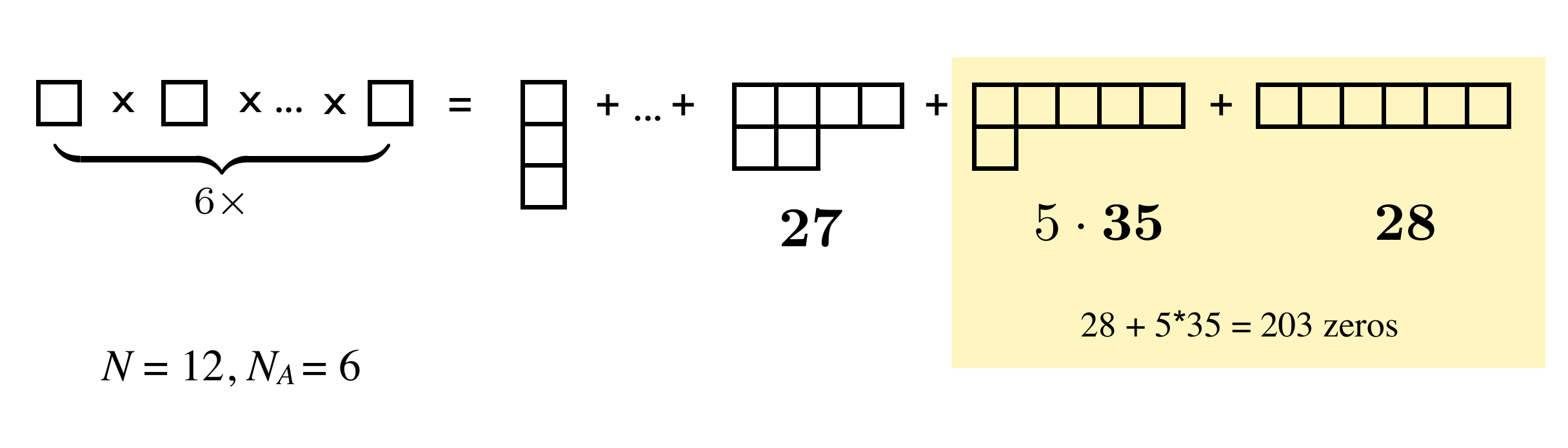}
\caption{Young tableaux decomposition of all SU(3) representations contained in the $N_A=6$ Schmidt basis. The representations that do not comply with the SU(3) bipartition sum rule for an SU(3) singlet ground state are highlighted in orange, corresponding to $\ge 5$ symmetrized blocks.}
\label{fig:general}
\end{figure}

 A notable aspect of $\rho\nd_A$ for finite size realizations of the ULS point is the observation of an extensive number of zeroes which are due to the enhanced SU(3) symmetry. To illustrate the latter for $N=12$, note first that because of its singlet property, any ground state configuration should contain 4 green, blue, and red particles according to the notation in Fig.~\ref{fig:su3-su3representations}, which directly connects to 4 sites located in the state $S_i^z=-1$, $0$, and $1$, respectively. Considering now the Schmidt basis of region $A$ (Fig.~\ref{fig:cut}b), it trivially follows that any basis configuration will contribute a zero eigenvalue which would necessarily yield a total state configuration violating the SU(3) sum rule mentioned before. For $N=12$ and $N_A=6$, this corresponds to all configurations with $5$ or $6$ particles of one color in the Schmidt basis. Since all configurations form SU(3) multiplets, such configurations are interpreted as the seed states for the multiplets they belong to which are all disallowed, i.e. the total number of zeroes is given by the total dimensionality of the associated SU(3) representations. How many such representations, and hence how many zeroes of $\rho\nd_A$, there are can be elegantly expressed in terms of SU(3) Young tableaux. For $N=3*n$, $n\in \mathbb{N}$, it corresponds to counting all representations with a number of symmetrized blocks that exceed $n$ (Fig.~\ref{fig:general}). In our case, we have to consider $\bss{3}^{\otimes 6}=5\cdot\bss{1} \oplus 16\cdot\bss{8}\oplus 10\cdot\bss{10} \oplus 5\cdot\overline{\bss{10}}\oplus 9\cdot\bss{27}\oplus\bss{28}\oplus 5\cdot \bss{35}$, and
 the total number of zeroes hence gives $5\times 35+28=203$ in the ground state bipartition $\rho\nd_A$ due to SU(3) symmetry. Away from the ULS point, the zero modes of $\rho\nd_A$ for ULS become very small eigenvalues of $\rho\nd_A$ and hence yield entanglement levels at high $\xi$. It gives a generic explanation for the large amount of high entanglement energy levels in the quadrupolar phase. The argument of bipartition-induced zero mode representations carries over to analogous scenarios for SU($N$) entanglement spectra.

%The phase in $\pi/4 < \theta < \pi/2$ has been investigated numerically already early on~\cite{fath1991} and was concluded to be gapless with soft modes at $k=0,2\pi/3$ and $4\pi/3$, where the dominant correlations are found to be of quadrupolar type~\cite{itoi,trebst}. integrable SU$(3)_1$ critical point in the columnar phase: Uimin-Lai-Sutherland point~\cite{uimin,lai,sutherland}. It is generally involved to extract the critical theory content such as central charge from the finite size scaling of the entanglement entropy~\cite{max}. 

%Following the generalized SU(N) Haldane gap criterion~\cite{confine}, the SU(3) rep. 3 Heisenberg model is gapless. Find critical points~\cite{rach1}.

\section{Dimerized phase: Translation symmetry breaking}
\label{sec:dimer}
\subsection{SU(3) dimerized point at $\theta=3\pi/2$}
\begin{figure}[t]
\centering
\includegraphics[width=0.99\columnwidth,clip]{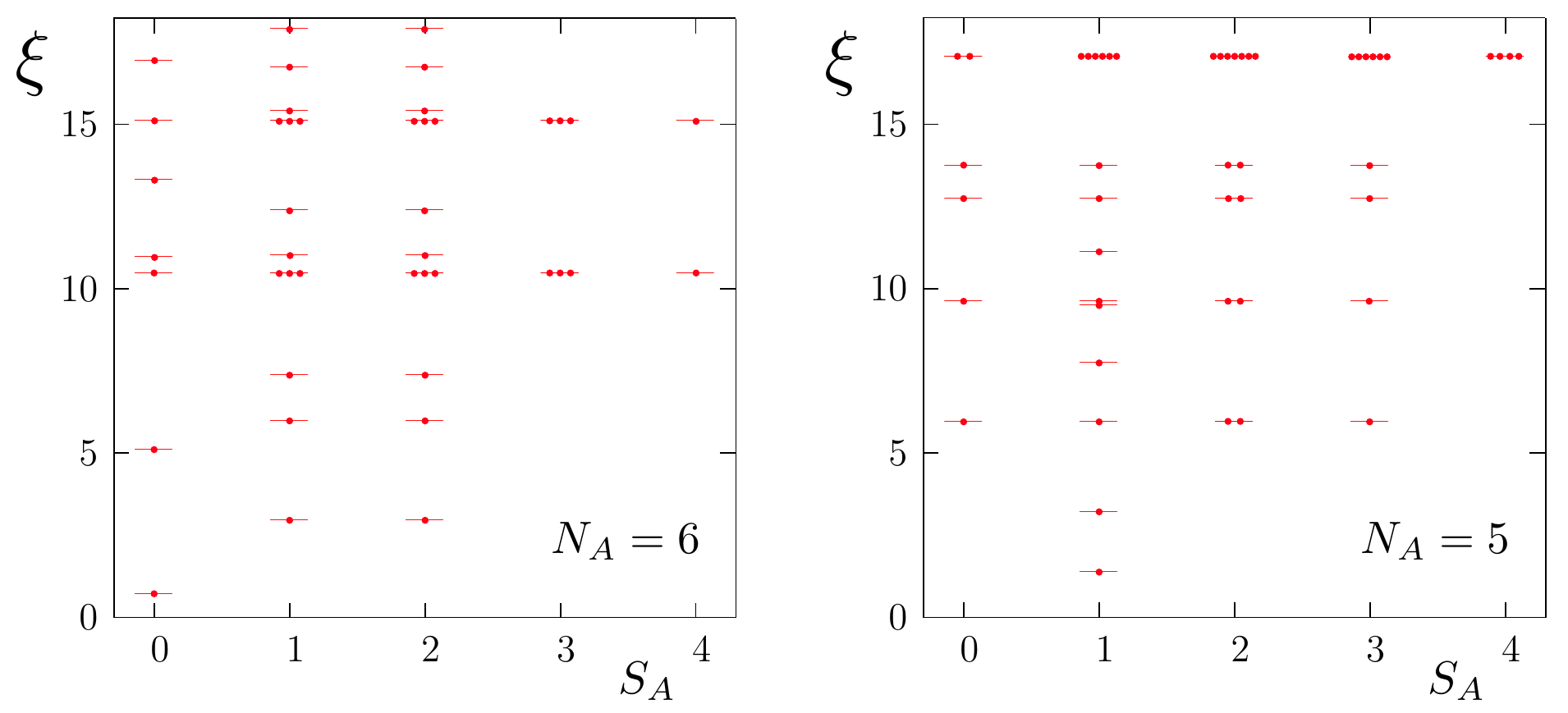}
\caption{$N=12$ entanglement cut for $N_A=5$ and $N_A=6$ for the dimerized SU(3) point at $\theta=3\pi/2$ for PBC. Aside from degeneracies between different $S_A$ sectors dictated by enhanced SU(3) symmetry, the low-lying entanglement spectrum of the gapped state switches between a singlet and triplet, which is reminiscent of translation symmetry breaking in the dimerized phase.}
\label{fig:dimersu3}
\end{figure}
%As alluded to  seen in Section~\ref{sec:quadru} and~\ref{app:lambda}
As alluded to in~\ref{app:lambda}, the model \eqref{ham} for $\theta=3\pi/2$ exhibits enhanced internal SU(3) symmetry similar to the ULS point discussed in Section~\ref{sec:quadru}. Here, however, the Hilbert space relates to an alternating $3$ and $\bar{3}$ representation. This has fundamental consequences on the very nature of the state. While it is still amenable to analytic solution, the state is not gapless but exhibits a gap along with translation symmetry breaking~\cite{barber-89prb4621,xian}, where the gap could be quantified by transfer matrix Bethe Ansatz~\cite{kluemper} and allows to estimate a spin-spin correlation length of 21 sites. It is revealing how the joint appearance of enhanced SU(3) and translational symmetry breaking manifests itself in the entanglement spectrum for small system sizes. To begin with, because of the $3\times \bar{3}$ Hilbert space structure, there is only an SU(3) sum rule for neighboring sites: in terms of SO(3) spin configurations for an even number of lattice sites, it implies that one finds non-zero weights in the ground state wave function only for basis states composed of two-site sequences $[S_i^z,S_{i+1}^z]=[1,-1],[-1,1],$ or $[0,0]$. As such, these configuration constraints only act locally on consecutive sites, and do not trigger extensive zero modes as observed in Section~\ref{sec:quadru}. 
%As such, the SU(3) symmetry of $\theta=3\pi/2$ is quantitatively not as characteristic for the dimerized phase as the ULS point is for the quadrupolar phase.

%{\it \underline{Philosophical remark:} in the same way the SU(3) symmetry with only fundamental representations does not allow for commensurate magnetic Neel order on a bipartite lattice, the entanglement matrix generates zeros for a bipartite cut. The $\bss{3}-\bar\bss{3}$ construction is, however, perfectly compatible with bipartite lattices and so is the entanglement matrix unaffected with bipartite cuts.}

 \begin{figure}[t]
\centering
\includegraphics[width=0.99\columnwidth,clip]{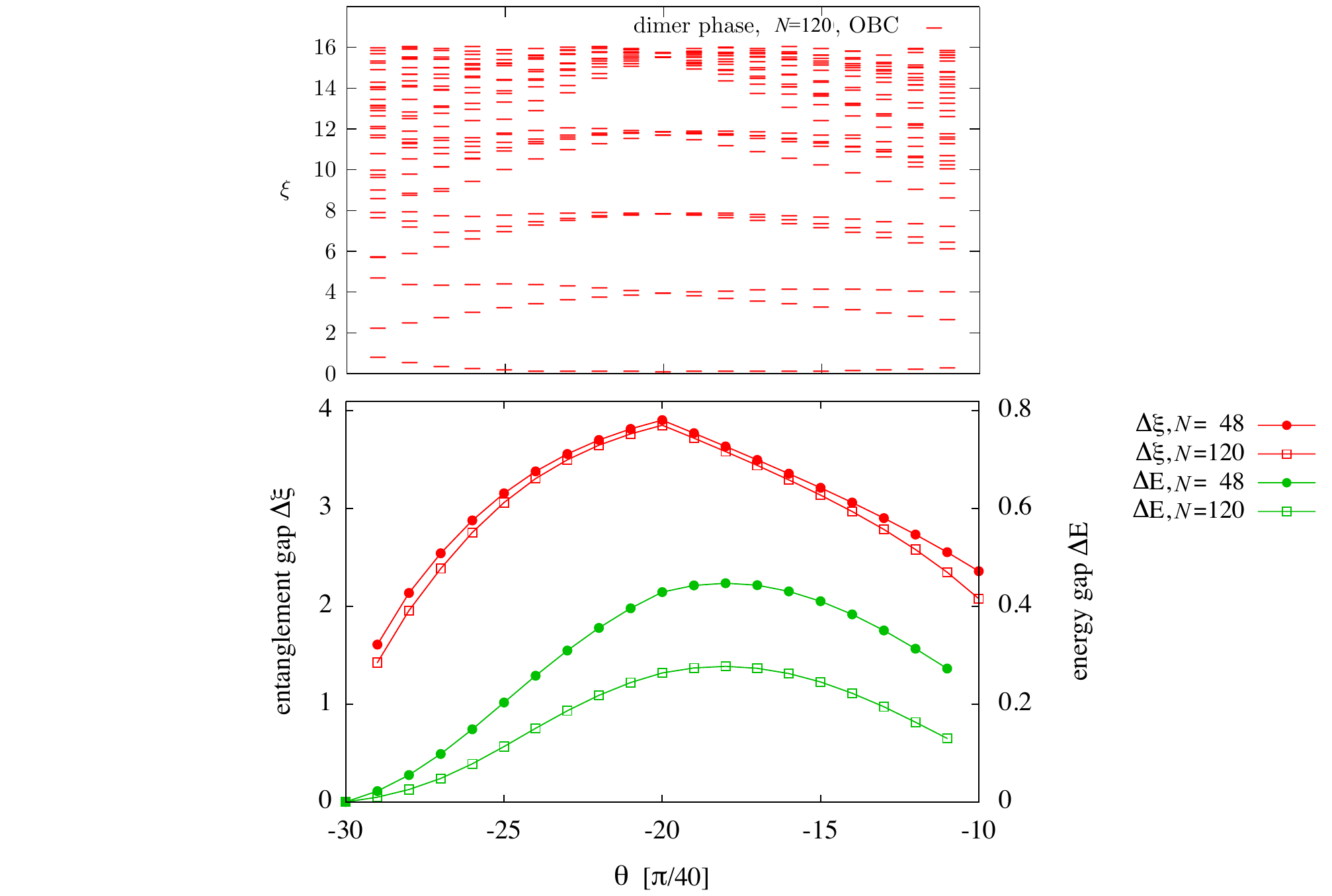}
\caption{Upper panel: spectral flow of entanglement spectra in the dimerized phase obtained within DMRG. Lower panel: energy gap (green) in units of~\eqref{ham} and entanglement gap (red) in units of Boltzmann temperature for $N=48$ and $N=120$ (OBC). The entanglement gap maximum is shifted against the energy gap maximum.}
\label{fig:dimer-phase}
\end{figure}

\subsection{Entanglement gap vs.\ energy gap}
The onset of translational symmetry breaking can be detected e.g. by analyzing $(N,N_A)=(12,5)$ and $(12,6)$ (Fig.~\ref{fig:dimersu3}). For gapped spin chains, the correspondence between the entanglement spectrum and the open boundary Hamiltonian spectrum can be developed (see \eg Ref.~\cite{alba}). This is because the low-energy modes contributing to the entanglement Hamiltonian, as located at the {\it Schmidt boundary}, correspond to the low-energy modes located at the {\it physical boundary}. For entanglement spectra in the dimerized phase, the lowest lying singlet state seen for $N_A=0$ mod $2$ alternates with the lowest lying triplet state for $N_A=1$ mod $2$. Note that this is not trivially dictated by the Hilbert space structure, as any even or odd tensor product of SO(3) representations contains singlets. Upon inspection, both low-energy spectra for $(N,N_A)=(12,5)$ and $(12,6)$, in terms of spectral structure and eigenvectors, correspond to an OBC $N=5$ and $N=6$ realization of~\eqref{ham}. The even chain yields a lowest-lying dimerized singlet, while the lowest-lying odd chain state breaks a dimer singlet into a triplet state.

This spectral structure carries over to the entanglement spectrum of the whole dimer phase. For the $N_A=6$ cut, an entanglement gap between the singlet state and the second lowest quintuplet state ($\theta> 3\pi/2$) or triplet state ($\theta < 3\pi/2$) can be defined ({\it cf}.\ Fig.~\ref{fig:dimer-phase}), with a corresponding level crossing cusp at $\theta=3\pi/2$~\cite{orus} (see also~\ref{app:egap-scaling}). Interestingly, the energy gap shows a similar, but not identical behavior. As can be also supported by larger scale density matrix renormalization group (DMRG) calculations, the energy gap maximum does not match the entanglement gap maximum (Fig.~\ref{fig:dimer-phase}). It stresses once again that while certain universality features are shared between energy spectra and entanglement spectra in gapped phases, this does not extend to non-universal features such as gap maxima.

\section{Symmetry-protected topological Haldane phase}
\label{sec:haldane}

\begin{figure}[t]
\centering
\includegraphics[width=0.99\columnwidth,clip]{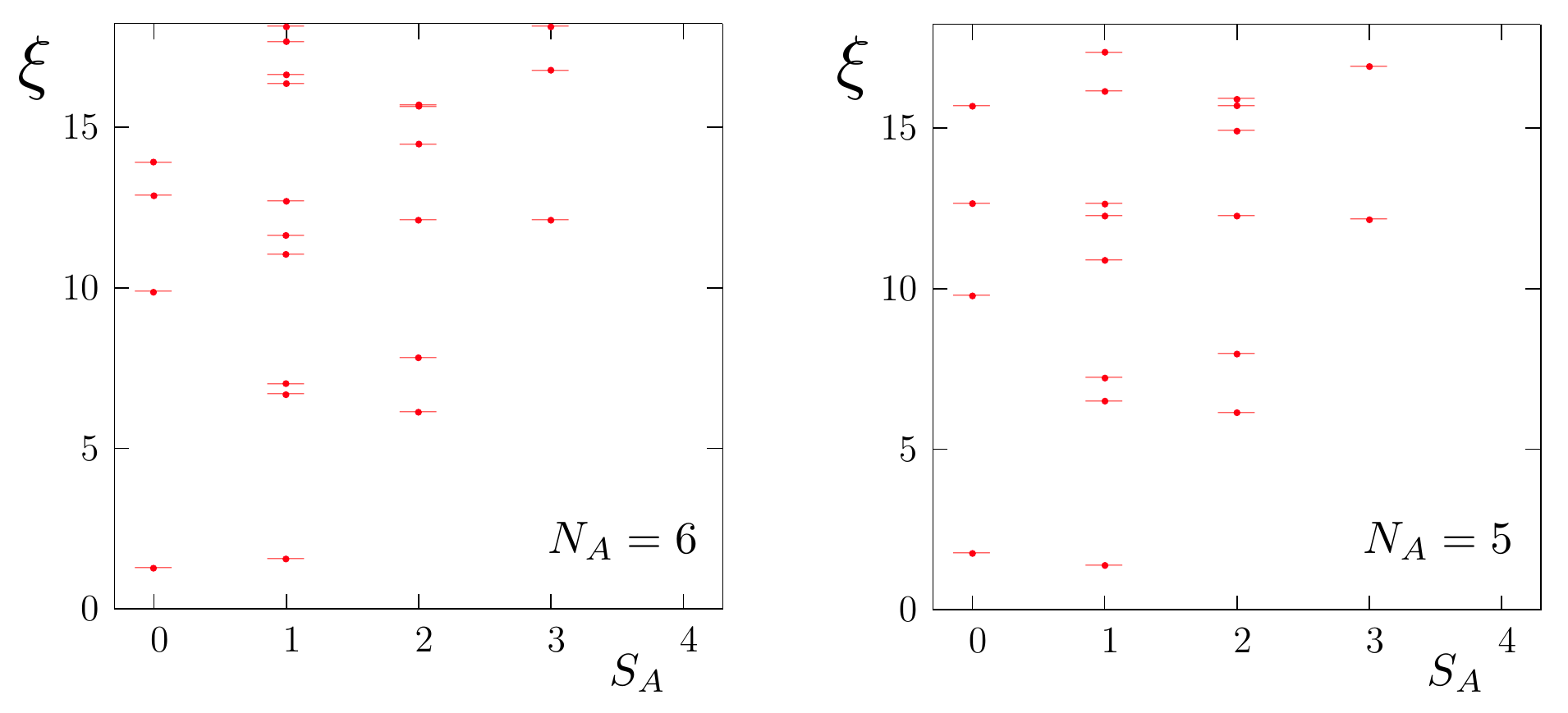}
\caption{Entanglement spectrum for $(N,N_A)=(12,6)$ and $(12,5)$ in the Haldane phase at $\theta=\pi/20$ for PBC. The low energy spectrum is given by a triplet and a singlet separated from higher levels. The spectrum is similar to the corresponding Hamiltonian spectrum for open boundaries.}
\label{fig:haldane-vs-dimer}
\end{figure}

The Haldane phase exhibits an energy gap without breaking of translation symmetry. The entanglement analysis shows no breaking of any other symmetry of~\eqref{ham}. The correspondence between the entanglement spectrum and the OBC Hamiltonian spectrum can be studied already for small finite system size. Fig.~\ref{fig:haldane-vs-dimer} shows the entanglement spectrum for even and odd $N_A$. The general structure of the entanglement spectrum stays unchanged, in clear contrast to the analogous analysis for the dimerized phase in Fig.~\ref{fig:dimersu3}. A triplet and a singlet level appear to form a separated low-energy set from the rest of the energy spectrum, where the singlet (triplet) state is the lowest entanglement energy state for even (odd) $N_A$. This is identically found for the OBC Hamiltonian spectrum.

\subsection{Entanglement gap vs.\ energy gap}
\begin{figure}[t]
\centering
\includegraphics[width=0.99\columnwidth,clip]{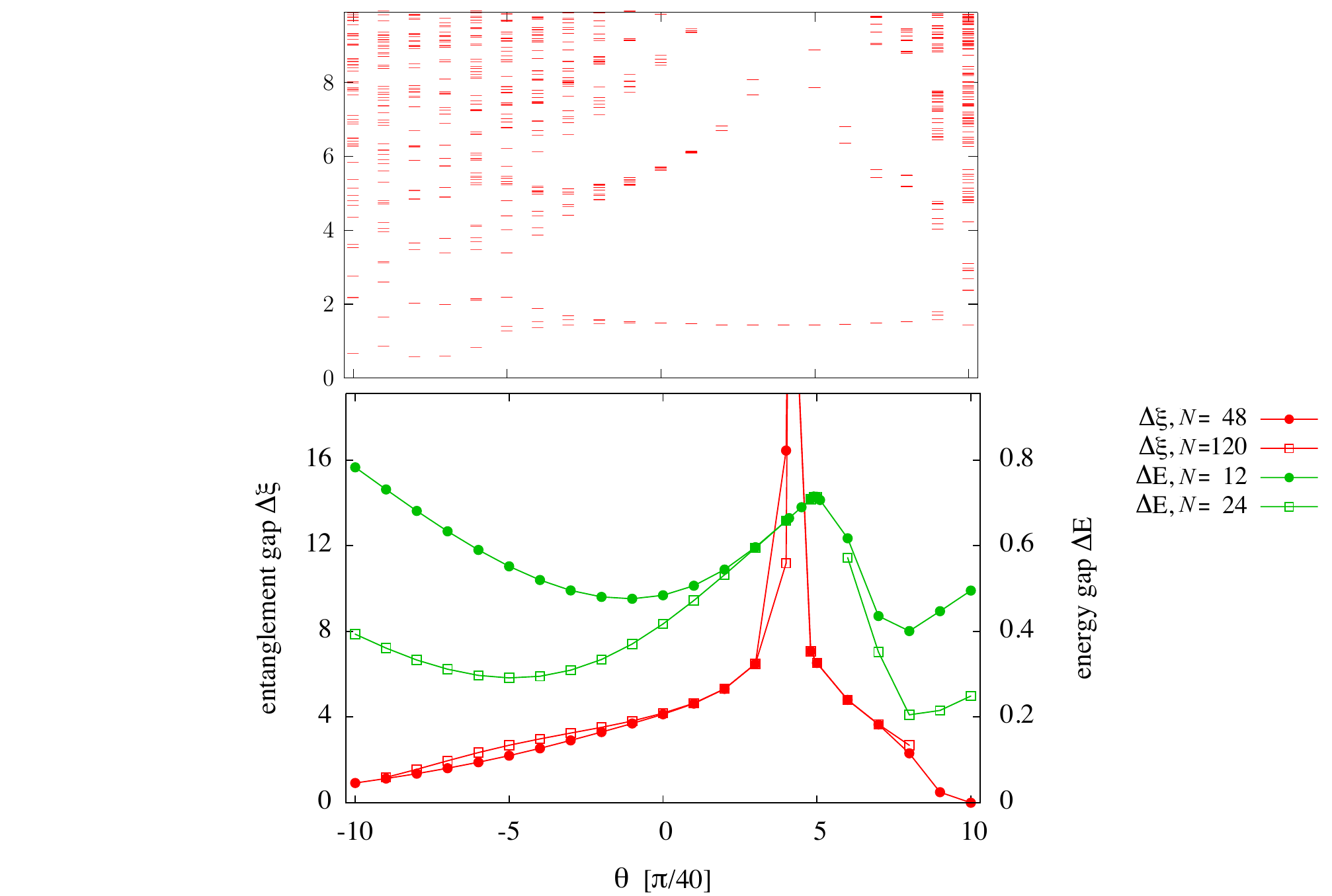}
\caption{Upper panel: Entanglement spectral flow of the Haldane phase obtained within DMRG for $N=48$ (PBC). Lower panel: Energy gap vs.\ entanglement gap for OBC. The proper investigation of the energy gap for large system sizes yields gap closure at $\pm \pi/4$ and a maximum energy-gap position shifted against the AKLT point~\cite{schollwoeck}.}
\label{fig:haldane-phase}
\end{figure}

The gap of the Haldane phase, and its entanglement Hamiltonian correspondence to OBC spectra derived from there, is identified as a Berry phase effect in integer spin chains~\cite{haldane-phase} and characterized by the AKLT point where the gap can be calculated analytically~\cite{aklt}. The AKLT model has become the paradigmatic symmetry-protected topological phase in one dimension~\cite{franky}, where the OBC low-energy behavior is characterized by one dangling spin-${1\over 2}$ degree of freedom at each boundary.   This likewise characterizes the low-energy entanglement spectrum, \ie the dangling spins form a triplet and a singlet which are separated from the other levels by an entanglement gap or energy gap, respectively (\ref{app:egap-scaling}). While the energy gap always stays finite, the entanglement gap becomes infinite at the AKLT point, where the ground state can be exactly characterized by a matrix product state with bond dimension $\mathcal{D}=2$. 

Fig.~\ref{fig:haldane-phase} depicts the entanglement spectral flow through the Haldane phase along with a comparison of energy gap vs.\ entanglement gap. From the ALKT point, branches of levels come down in entanglement energy and eventually close the entanglement gap towards the ULS point at $\pi/4$ and the Takhtajan-Babujian point at $-\pi/4$, respectively. The energy gap similarly closes at $\pm \pi/4$ as seen in finite size scaling, but exhibits a gap maximum shifted against the entanglement gap maximum at the AKLT point. (The observation from small finite size is confirmed by large scale calculations, see e.g. Ref.~\cite{schollwoeck}.) This is similar to the dimerized phase in Fig.~\ref{fig:dimer-phase}.

%The same phenomena can be observed in the Haldane phase. Very brief discussion. 
%Next discuss the AKLT point and its peculiarities. Discuss OBC vs.\ PBC cut here?

%%%%%%%%%%%%%%%%%%%%%%%%%%%%%%%%%%%%%%%%%%%%
%
%           S  M  A
%
%%%%%%%%%%%%%%%%%%%%%%%%%%%%%%%%%%%%%%%%%%%%
\subsection{Single mode approximation}
\label{subsection:sma|}
Our starting point is the AKLT Hamiltonian plus the Heisenberg term as a small perturbation,
\begin{equation}
H=H_{\rm AKLT} + \eps \sum_i \bss{S}_i\cdot \bss{S}_{i+1}\ .
\label{AKLTpert}
\end{equation}
Using the spectrum of single mode approximation (SMA) excitations, it is possible to derive an approximate perturbation theory
in the difference $H-H_{\rm AKLT}$. (Note that SMA was employed previously to elucidate the relation between valence bond states (VBS) and the Laughlin state~\cite{dan}.)
The idea goes back to unpublished work of Haldane, and was later elucidated in Ref. \cite{DPA-SMG-92}.
One proceeds by idealizing the triplet SMA levels as bosonic excitations, and asserting the approximate operator correspondence
\begin{equation}
S^\alpha_k\approx\sqrt{s(k)}\,\big(b\yd_{k,\alpha}+b\nd_{-k,\alpha}\big),
\end{equation}
where $s(k)=\langle\Psi_0|S^\alpha_k S^\alpha_{-k}|\Psi_0\rangle$ (no sum on $\alpha$) is the static structure factor.
The bosonic form of the perturbed Hamiltonian in~\eqref{AKLTpert} is then
\begin{equation}
H_{\rm bos}=\sum_{k,\alpha}\Big[ \omega(k) b\yd_{k,\alpha} b\nd_{k,\alpha} + \varepsilon  s(k)\cos k
\big(b\yd_{k,\alpha}+b\nd_{-k,\alpha}\big)\big(b\yd_{-k,\alpha}+b\nd_{k,\alpha}\big)\Big],
\end{equation}
where the structure factor and SMA dispersion were computed in Ref. \cite{dan} and found to be $s(k)=2(1-\cos k)/(5+3\cos k)$ and
$\omega(k)=\frac{5}{27}(5+3\cos k)$.  Solving $H_{\rm bos}$ by a Bogoliubov transformation, one obtains the renormalized structure factor
\begin{equation}
{\tilde s}(k)=\left( \frac{\omega(k)}{\omega(k) + 2\eps\Delta(k)}\right)^{\!1/2} s(k)\ ,
\end{equation}
where $\Delta(k)=2\,s(k) \cos{k}$. The corresponding spin wavefunction is then given by
\begin{equation}
\ket{{\widetilde\Psi}_0}=\exp\!{\left( - \sum_k\left( \frac{1}{{\tilde s}(k)} - \frac{1}{s(k)} \right) S^\alpha_k S^\alpha_{-k} \right)} \ket{\Psi_0}\ ,
\end{equation}
where $\ket{\Psi_0}$ denotes the AKLT ground state.
Expanding to leading power in $\eps$, we have
\be
\ket{{\widetilde\Psi}_0}= \left\{ 1 -\frac{\eps}{2}\! \sum_{n,n'} K_{nn'} \,S^\alpha_n S^\alpha_{n'} + {\cal O}(\eps^2)\right\}\ket{\Psi_0}\ ,\label{thirdeqn}
\ee
where the kernel $K_{nn'}$ is the Fourier transform of $2\Delta(k)/\omega(k) s(k)=4\cos k/\omega(k)$, and is given by the expression ($j:= n-n' $)
\be
K(j)=\int\limits_0^{2\pi}\!{d k\over 2\pi}\>{4\cos k\over\omega(k)}\>e^{ikj}
=\frac{36}{5}\,\eps\,\delta_{j,0} -9\,\eps\,\bigg(\!\!-\frac{1}{3}\bigg)^{j}\ .
\label{Kker}
\ee
The expression for $\ket{{\widetilde\Psi}_0}$ has a clear correspondence with first order perturbation theory, wherein
\begin{equation}
\ket{\Psi'_0}= \ket{\Psi_0} - \sum_n {1\over E_n} \ket{n} \!\bra{n} 2\eps\!\sum_{k} \cos(k)\,S^\alpha_k S^\alpha_{-k} \ket{\Psi_0}\ ,
\end{equation}
if we approximate $\ket{n}\approx s(k)^{-1}\,\big(1-\ket{\Psi_0}\bra{\Psi_0}\big) \, S^\alpha_k S^\alpha_{-k}\,\ket{\Psi_0}$,
and $E_n\approx \omega(k)$.

\subsection{Operator product expansion of the SMA wave function}
Defining the operator-valued matrix,
\begin{equation}
M(j)= \left( \begin{array}{ccccc}
1 & S^x_j & S^y_j & S^z_j & 0 \\ 0 & e^{-\alpha} & 0 & 0 & S^x_j \\
0 & 0 & e^{-\alpha} & 0 & S^y_j \\  0 & 0 & 0 & e^{-\alpha} & S^z_j \\ 0 & 0 & 0 & 0 & 1
\end{array}\right)\ ,
\end{equation}
and the vectors
\begin{equation}
\bra{{\rm L}}=\left( \begin{array}{ccccc}-3\eps & 0 & 0 & 0 & 1\end{array}\right)\quad,\quad
\ket{{\rm R}}=\left( \begin{array}{c} 0 \\ 0 \\ 0 \\ 0 \\ 1 \end{array}\right)\ ,
\end{equation}
we have the matrix product operator (MPO) expression,
\be
\bra{{\rm L}}M(1)\cdots M(N)\ket{{\rm R}}=1+\sum_{n < n'} K(n-n')\,S^\alpha_n S^\alpha_{n'}\ .
\ee

If the AKLT state is written in matrix product form as 
\begin{equation}
\ket{\Psi_0}={\bra{\cal L}}{\cal A}^{m_1}\cdots{\cal A}^{m_N}\ket{{\cal R}}\,\ket{m_1,\ldots,m_N}\ ,
\end{equation}
where $\bra{{\cal L}}$ and $\bra{{\cal R}}$ are vectors which passivate the end sites $n=1$ and $n=N$, rendering them $S={1\over 2}$,
then the SMA state is a MPS with
\begin{equation}
\ket{\Psi}=\bra{L}{A^{m_1}\cdots A^{m_N}}\ket{R}\,\ket{m_1,\ldots,m_N}\quad,
\end{equation}
where
\begin{equation}
A^m_{i,i'}=\bra{m}{M_{aa'}}\ket{m'}\,{\cal A}^{m'}_{\mu,\mu'}\ ,
\end{equation}
and $i=(a,\mu)$ is a composite index.  In our example, $a$ runs from $1$ to $5$, $\mu$ from $1$ to $2$, and $m$ from $1$ to $3$.
Thus, $i$ runs from $1$ to $5\cdot 2=10$.  Similarly, $\ket{L}=\ket{{\rm L}}\otimes\ket{{\cal L}}$, {\it etc.}
The fact that $A^m_{i,i'}$ is of rank ten means that there must be ten entanglement levels.

\begin{figure}[t]
\centering
\includegraphics[width=0.95\linewidth]{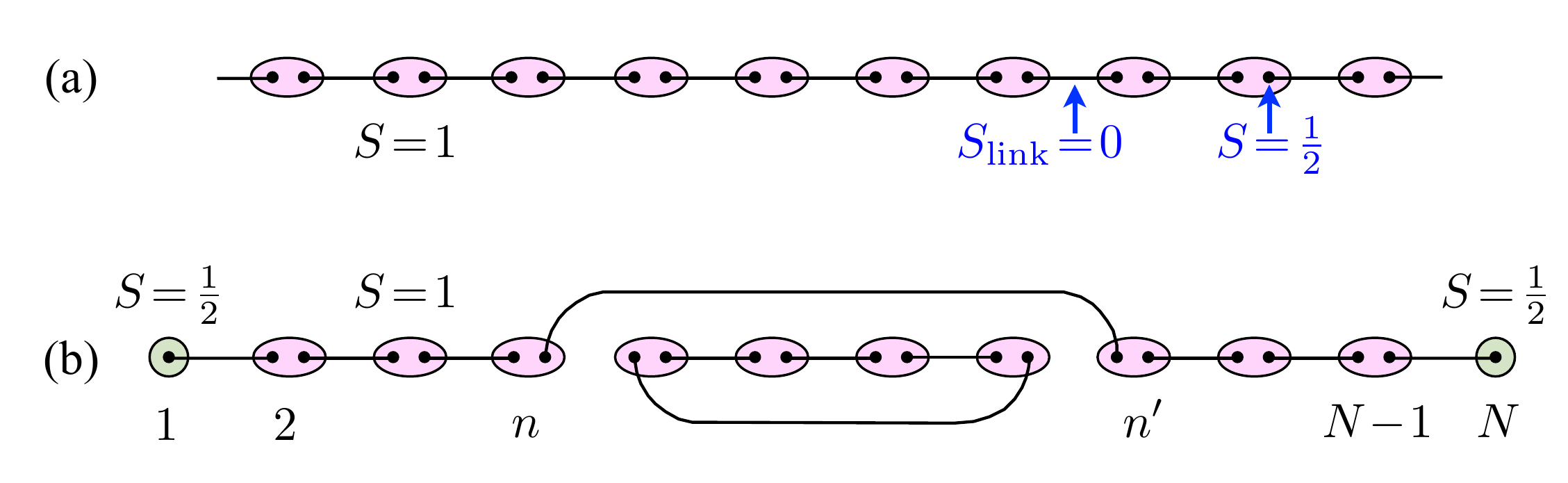}
\caption{(a) $S=1$ AKLT state; each lattice site (pink ellipses) is given by the symmetrization of two virtual spin-${1\over 2}$ degrees of freedom. One of them is antisymmetrized in the singlet bond to the left while the other is antisymmetrized in a singlet bond to the right. (b) Visualization of the state $\ket{nn'}$ within the SMA framework for the AKLT chain with passivated $S={1\over 2}$ ends.}
\label{fig:nnprime}
\end{figure}

More frequent than a double cut of a finite size bipartition, the single cut associated with a semi-infinite partition has been employed to identify the symmetry-protected topological character of the Haldane phase~\cite{franky}. The latter is simulated for small finite size by passivating the dangling spins in an OBC geometry (Fig.~\ref{fig:cut}b) and choosing the $S^z_{\rm tot}=\pm 1$ sector. Combining the valence-bond picture (see Fig.\,\ref{fig:nnprime}) and the framework of SMA, we will demonstrate that the ES in the Haldane phase must be two-fold (or, in general, even-numbered) degenerate. 

%%%%%%%%%%%%%%%%%%%%%%%%%%%%%%%%%%%%%%%%%
%%%%%%%%%%%%%%%%%%%%%%%%%%%%%%%%%%%%%%%%%%

Suppose we partition the chain such that sites $j \in \{1, 2,...,N_{\rm A}\}$ are in subsystem A and the remaining sites are in subsystem B. It follows that we may write
\begin{equation}
\ket{\Psi_{\rm SMA}} = \ket{\Psi_0} + \ket{\Psi_{\rm A}} + \ket{\Psi_{\rm B}} + \ket{\Psi_{\rm AB}} ,
\end{equation}
where $\ket{\Psi_{\rm A}}$ includes contributions for $n < n' \leq N_{\rm A}$, $\ket{\Psi_{\rm B}}$ includes contributions for $N_{\rm A} < n < n'$, and $\ket{\Psi_{\rm AB}}$
includes contributions for $n < N_{\rm A} < n'$ (Eq. \ref{thirdeqn}).

We may then write the density matrix $\rho = \ket{\Psi}\bra{\Psi}$ as $\rho = \rho\nd_0 + \rho\nd_{\rm A} + \rho\nd_{\rm B} + \rho\nd_{\rm AB}$, where
$\rho\nd_{\rm A}=\ket{\Psi_0}\bra{\Psi_{\rm A}} + \ket{\Psi_{\rm A}}\bra{\Psi_0}$, $\rho\nd_{\rm B} = \ket{\Psi_0}\bra{\Psi_{\rm B}} + \ket{\Psi_{\rm B}}\bra{\Psi_0}$, and
$\rho\nd_{\rm AB} = \ket{\Psi_0}\bra{\Psi_{\rm AB}} + \ket{\Psi_{\rm AB}}\bra{\Psi_0}$.
We now perform the partial trace over the B subsystem. For $\rho\nd_0$, $\rho\nd_{\rm A}$, and $\rho\nd_{\rm B}$, this results in one link being severed by the entanglement cut.
There are then exactly two entanglement eigenstates in each case (\ie a doublet).  For $\rho\nd_{\rm AB}$, three links are severed (see Fig.\,\ref{fig:nnprime}), resulting in two doublets
and one quadruplet entanglement levels. Assuming that the doublets are linearly independent, we find three linearly independent doublets and
one quadruplet, resulting in ten entanglement levels, as previously deduced.
The upshot of this analysis is that one predicts the appearance of doublets and quadruplets, but nothing else.  Here we recover the two-fold degeneracy (or, more generally,
even-number degeneracy) of the ES for the Haldane phase~\cite{franky}. If we allow for higher processes and go beyond the linearized SMA version considered here, we will generate
higher-order links as compared to $\ket{nn'}$. Still only configurations are generated where the real space cut intersects an {\it odd} number of links. Therefore the ES will remain
even-numbered degenerate.

\begin{figure}[t]
\centering
\includegraphics[width=0.60\columnwidth,clip]{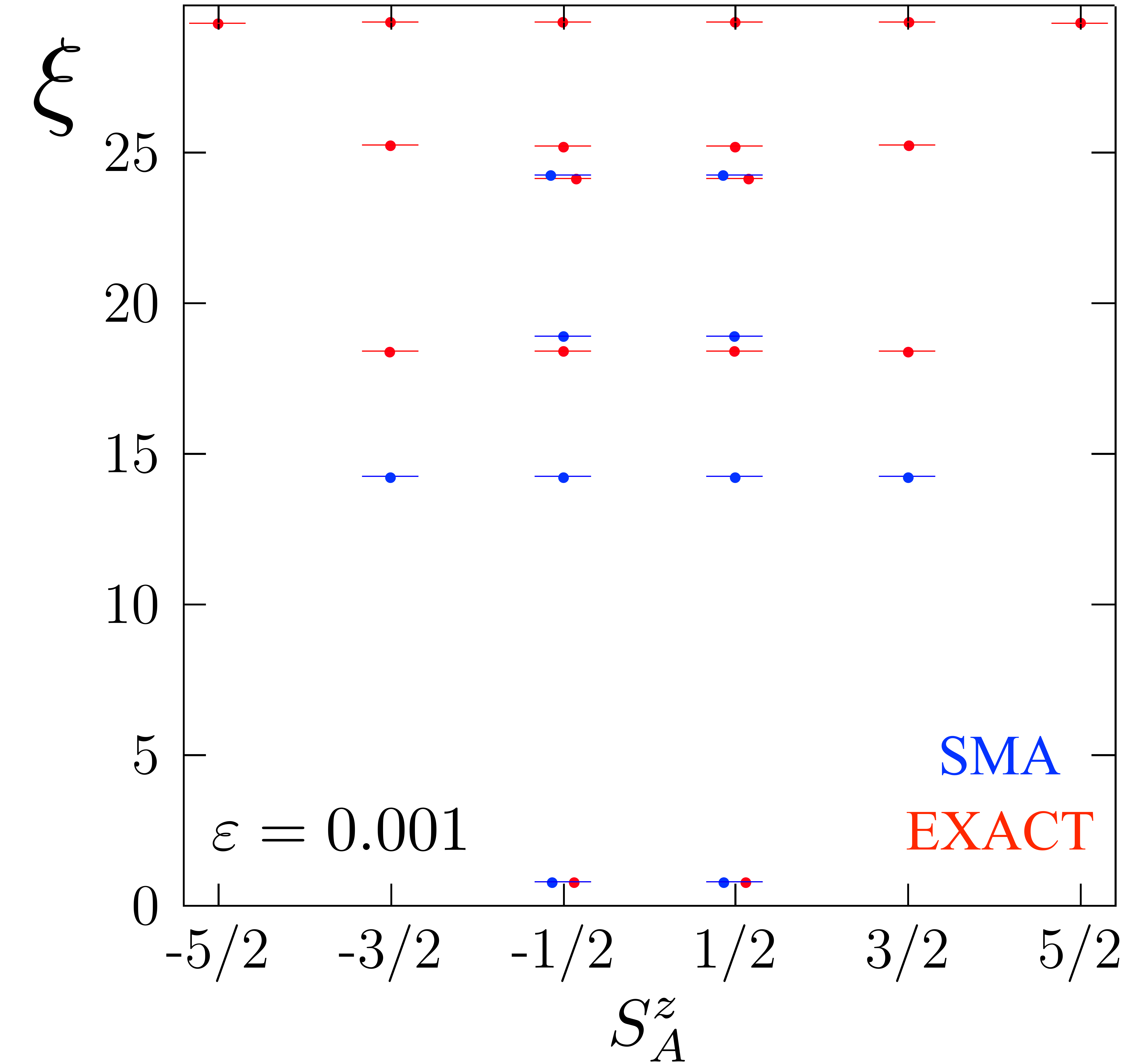} %es_SMA}
\caption{Comparison of the SMA-corrected ground state entanglement spectrum (leading order in $\eps$, blue) (Eq.~\ref{thirdeqn}) and the exact ground state entanglement spectrum for a single cut in a $N=12$ site chain of the Hamiltonian in Eq.~\ref{AKLTpert} (red).
All entanglement levels are even-fold degenerate in both cases.  Since we mimic the single cut by imposing OBC and $S^z_{\rm tot}=1$, exact SU(2) symmetry in the ES is
lost, but is still approximately present.}
\label{fig:sma}
\end{figure}

The double degeneracy of the ES for a single cut in the Haldane phase clearly distinguishes the Haldane phase from the dimer phase.
While the SMA picture can explain the two-fold degeneracy in the ES, one might naively expect that breaking of SU(2) symmetry would lift these degeneracies. This is, however, not true. As shown in Ref.\,\cite{franky}, the Haldane phase is {\it protected} by time-reversal, bond-inversion, and dihedral symmetries. One needs to break all these symmetries in order to loose the degeneracy of the entanglement spectrum. In the meantime it is well-established that the Haldane phase is an SPT phase\,\cite{XGWen_Science2012}. The double degeneracy in the entanglement spectrum is a hallmark for such a SPT phase when only a single cut is considered, as just explained. When PBCs are imposed and two cuts are present, multiplets stemming from the two cuts decompose into multiplets such as singlet, triplet etc. The simplest example is the pure AKLT state where a doublet on both cuts is present resulting in one singlet and one triplet level. There is no non-trivial even/odd degeneracy constraint anymore.

%\begin{figure}[t]
%\centering
%\includegraphics[width=0.95\columnwidth,clip]{exact_vs_sma_smalleps} 
%\caption{Entanglement spectra in a $L=12$ site $S=1$ ring with $H=H_{\rm AKLT}+\eps H_{\rm He is}$, cut into two six-site half-chains.  The approximate fourfold
%degeneracy of the lowest-lying entanglement levels is due to the two $S={1\over 2}$ entanglement eigenstates associated with each single cut.  Magnifications show structure
%of higher-lying entanglement levels.}
%\label{fig:smab}
%\end{figure}

\begin{figure}[t]
\centering
\includegraphics[width=0.75\columnwidth,clip]{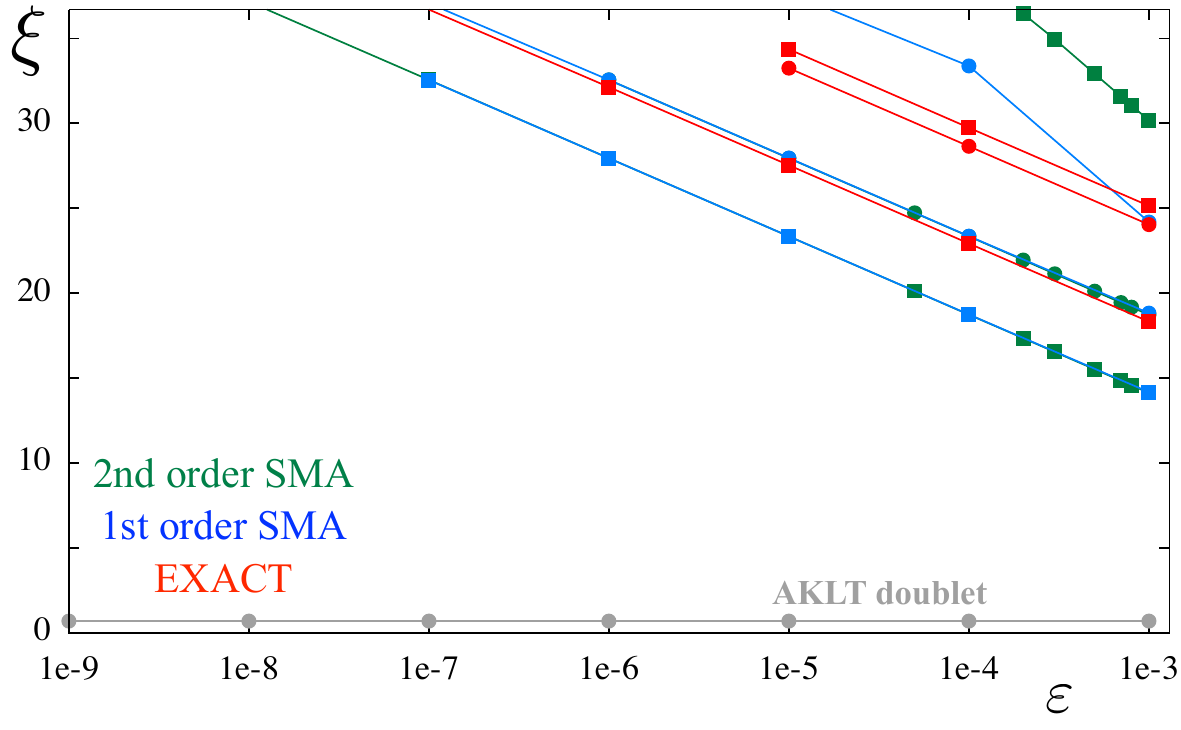} 
\caption{Evolution of entanglement spectra for an $N=12$ site chain of Eq.\,\eqref{AKLTpert} as a function of $\eps$.  A single cut is used with $S^z_{\rm tot}=1$
enforced to passivate the two end spins.  Comparisons of the exact ground state (red) with first (blue) and second (green) order SMA-corrected ground states are shown. All states exhibit the characteristic AKLT doublet whose  $\xi$ level does not change with $\eps$ within machine precision. All doublets are depicted by full circles and the quartets by full squares. The lowest entanglement levels above the AKLT doublet are hardly affected by moving from 1st to 2nd order SMA. The quantitative deviation of the SMA levels from the exact levels is small in terms of absolute entanglement weight, given the high values of $\xi$.}
\label{fig:smac}
\end{figure}

\subsection{Comparison of SMA predictions with numerical results}
We test the SMA method quantitatively by comparing the entanglement spectrum of the SMA-modified ground state wavefunction with that of the exact ground state.
We first consider an open chain with $N=12$ sites and make a single entanglement cut which divides the system into two half-chains.  The end spins are passivated by
fixing $S^z_{\rm tot}=1$.  In Fig.\,\ref{fig:sma} we show both SMA (blue) and exact ES (red) for $\eps=0.001$.  The lowest-lying doublet, the entanglement ground state,
is followed by a quadruplet and another doublet.  The SMA state entanglement level multiplet structure is as predicted above ({\it i.e.} ten levels in total), but deviates
noticeably from the exact result for entanglement levels above the lowest doublet.  One reason is that the SMA-corrected ground state is of the form
$\exp\left(\sum_{j=1}^\infty \eps^j\,{\widehat{\cal Q}_j}\right)\ket{\Psi_0}$, where $\{\widehat{\cal Q}_j\}$ are operators which scale extensively with system size. 
Hence an expansion of the exponential is really an expansion in powers of $N\eps$.
We can do a little better by expanding the SMA-corrected ground state wavefunction to second order, writing
\be
\ket{{\widetilde\Psi}_0}=\left\{ 1 - {\eps\over 2}\,{\cal M}^{(1)} + {\eps^2\over 8}\,{\cal M}^{(2)}+ {\eps^2\over 8}\left({\cal M}^{(1)}\right)^2+\mathcal{O}(\eps^3)
\right\}\ket{\Psi_0}\ ,
\ee
where ${\cal M}^{(r)}=\sum_k \frac{1}{s(k)} \left(\frac{2 s(k) \cos k}{\omega(k)} \right)^r S_k^\alpha S_{-k}^\alpha=\sum_{n,n'} K^{(r)}_{n-n'}\,S^\alpha_n\,S^\alpha_{n'}$, with $K^{(1)}_{j}\equiv K(j)$ in Eq. \ref{Kker} above, and
\be
K^{(2)}_j=-\frac{864}{25}\,\delta\nd_{j,0} + \frac{27}{200}\left(-\frac{1}{3}\right)^{j}\!\!\cdot (527-180 j + 400 j^2)\ .
\ee
In Fig. \ref{fig:smac} we compare the low-lying entanglement levels related to Eq.~\ref{AKLTpert} for the exact ground state
as well as the first and second order SMA results, as a function of $\eps$.  We find that the ES is strongly affected for levels above the lowest eight states arranged in the
doublet, quadruplet, doublet order. Due to the second order contribution, the sequence of entanglement levels (d=doublet and q=quartet)  changes from d-q-d-d for the first order SMA to d-q-d-q, which matches the exact sequence. Still, the second order contribution modifies the entanglement profile only marginally.
%We have also computed the ES in a $L=12$ site ring, dividing it with two cuts into two six-site chains.  The entanglement ground
%state is then approximately fourfold degenerate, due to the two $S={1\over 2}$ entanglement states localized along each cut.  
The fact that there is still only a rough quantitative correspondence in $\xi$
between the higher lying entanglement levels for the SMA-corrected and exact ground state indicates that the source of the disagreement lies in the
approximate nature of the Bogoliubov-SMA approach itself, rather than the passivation of the end spins by projecting onto $S^z_{\rm tot}=1$, or the finite order expansion in $N\eps$.

\section{CONCLUSION}
\label{sec:conclusion}
We have employed finite size ground state bipartitions to analyze the entanglement structure of isotropic spin-1 chains. The isotropic bilinear-biquadratic model allowed us to interpret many hallmark features of quantum spin chains from the viewpoint of entanglement.  We investigated spontaneous breaking of translation symmetry in the dimerized phase and spin rotation symmetry in the ferromagnetic domain. Both phenomena can be unambiguously identified from entanglement just by analyzing a single small system size. Furthermore, we have shown how enhanced internal SU(3) symmetry affects not only the degeneracy structure of entanglement spectra similar to Hamiltonian spectra, but also the pronounced extensive amount of zero weight entanglement levels as seen for the Uimin-Lai-Sutherland point. This spectral distribution feature of entanglement approximately persists for the whole quadrupolar phase. Finally, the SPT character of the Haldane phase is precisely resolved by entanglement spectra. We have elucidated the correspondence between a boundary Hamiltonian spectrum and the entanglement spectrum of the associated ground state for periodic boundary conditions. In particular, we have identified the notion of the entanglement gap to hold in the thermodynamic limit. For the single mode approximation of the perturbed AKLT state and the operator product expansion derived form there, entanglement spectra allowed us to obtain a complementary view on the accuracy of this approach. 

Our study strongly supports the view that the analysis of entanglement spectra of one-dimensional quantum systems provides insight which leverages the analysis of relatively small
finite size systems, and will thereby constitute a preferable choice when large scale calculations are either not feasible or unavailable.

\section*{Acknowledgments}
We thank F. Pollmann, P. Schmitteckert, and U. Schollw\"ock for valuable discussions. This work has been supported by the ERC Starting Grant ERC-StG-Thomale-TOPOLECTRICS-336012. SR is supported by the DFG through FOR 960 and by the Helmholtz association through VI-521. BAB acknowledges support from NSF CAREER DMR-0952428, ONR-N00014-11-1-0635, MURI-130-6082, DARPA under SPAWAR Grant No.: N66001-11-1-4110, the Packard Foundation, and a Keck grant.

\appendix

\section{Derivation of enhanced SU(3) symmetry models}
\label{app:lambda}

For completeness, we elaborate on the points of~\eqref{ham} where the Spin 1 SO(3) model exhibits enhanced internal SU(3) symmetry. We follow the notation of~\cite{muetter}, and prove below that both Hamiltonians $\theta=\pi/4$ and $\theta=3\pi/2$ of~\eqref{ham} possess an enlarged SU(3) symmetry. Since both models exhibit SU(3) singlet ground states, the same argument of for SU(2) in Section~\ref{sec:model} applies and as such, the internal symmetry operator generating the block diagonal form of the entanglement spectrum is given by $[T^a,\rho\nd_A]=0$, $T^a=\sum_i T^a_i$, where $T^a$ are the generators in sl(3). This yields SU(3) multiplets in the entanglement spectrum.

\subsection{SU(3) Heisenberg model}
To prove the SU(3) invariance at the Hamiltonian level, we develop an {\it a posteriori} perspective and consider the $SU(3)$ Heisenberg Hamilton operator
\begin{equation}
H=\sum_i^N T_i^a T_{i+1}^a, \label{su3}
\end{equation}
with implicit summation over $a=1,\dots 8.$
 We label the
representation states of the fundamental representation rep. $3$ by the colors blue (b), red (r), and green (g)
(quarks). We will also treat the non-equivalent
representation $\bss{\bar{3}}$ later, \ie  where the states possess the complementary
colours yellow (y), cyan (c), and magenta (m) (anti-quarks). Labelling the Gell-Mann representation by $\lambda^a$ in order not
to confuse it with the generators $T$, the action on the fundamental
representation $3$ is given by 
\begin{eqnarray}
\lambda^1&=&\left(\begin{array}{ccc}0&1&0\\1&0&0\\0&0&0\end{array}
\right)\!\!,\quad
\lambda^2 =\left(\begin{array}{ccc}0&-i&0\\i&0&0\\0&0&0\end{array}
\right)\!\!,\quad
\lambda^3 =\left(\begin{array}{ccc}1&0&0\\0&-1&0\\0&0&0\end{array}
\right)\!\!,\nonumber \\
\lambda^4&=&\left(\begin{array}{ccc}0&0&1\\0&0&0\\1&0&0\end{array}
\right)\!\!,\quad
\lambda^5 =\left(\begin{array}{ccc}0&0&-i\\0&0&0\\i&0&0\end{array}
\right)\!\!,\quad
\lambda^6 =\left(\begin{array}{ccc}0&0&0\\0&0&1\\0&1&0\end{array}
\right)\!\!,\nonumber \\
\lambda^7&=&\left(\begin{array}{ccc}0&0&0\\0&0&-i\\0&i&0\end{array}
\right)\!\!,\quad
\lambda^8 =\frac{1}{\sqrt{3}}
\left(\begin{array}{ccc}1&0&0\\0&1&0\\0&0&-2\end{array}\right)\!\!.
%\label{eq:appsu3gellmannmatrices}
\end{eqnarray}
Their normalization is chosen to be
\begin{displaymath}
\mathrm{tr}\left(\lambda^a\lambda^b\right)=2\delta_{ab}.
%\label{eq:appsu3-gellmannnormalization}
\end{displaymath}
The Gell-Mann matrices form an orthogonal basis of sl$_3$, the Lie algebra of
SU(3), and satisfy the commutation relations
\begin{equation}
\comm{\lambda^a}{\lambda^b}=2i f^{abc}\lambda^c.
%\label{eq:appsu3commutatorforlambda}
\label{antisymm}
\end{equation}
The structure constants $f^{abc}$ are totally antisymmetric and obey Jacobi's
identity
\begin{equation}
f^{abc}f^{cde}+f^{bdc}f^{cae}+f^{dac}f^{cbe}=0.
\label{eq:appsu3-JI}
\end{equation} 
All non-vanishing structure constants are obtained by 
permutations of the indices from
\begin{eqnarray}
f^{123}&=&1 \nonumber \\
f^{147}&=&f^{246}=f^{257}=f^{345}=\frac{1}{2} \nonumber \\
f^{156}&=&f^{367}=-\frac{1}{2} \nonumber \\
f^{458}&=&f^{678}=\frac{\sqrt{3}}{2} \; .
\end{eqnarray}
The Gell-Mann matrices also close with respect to the
anticommutator:
\begin{equation}
\{\lambda_a, \lambda_b \}= \frac{4}{3} \delta_{ab} + d_{abc}
\lambda_c,
\label{sym}
\end{equation}
with the fully symmetric structure constants
\begin{eqnarray}
&&d^{118}=d^{228}=d^{338}=-d^{888}=\frac{2}{\sqrt{3}} \nonumber \\
&&d^{448}=d^{558}=d^{668}=d^{778}=-\frac{1}{\sqrt{3}} \nonumber \\
&&d^{146}=d^{157}=-d^{247}=d^{256}=d^{344}=d^{355}=-d^{366}=-d^{377}=1
\end{eqnarray}

\begin{figure}[t]
\begin{center}
\setlength{\unitlength}{12pt}
\begin{picture}(29,11)(1,0)
\put(13,10){\textbf{3}}
\put(2,5){\line(1,0){10}}
\put(7,0){\line(0,1){10}}
\put(12,4.3){$T^3$}
\put(7.2,9.6){$T^8$}
\put(7,1.5){\circle*{0.5}}
\put(6,1.5){g}
\put(9.93,6.75){\circle*{0.5}}
\put(10.43,6.75){b}
\put(4.07,6.75){\circle*{0.5}}
\put(4.57,6.75){r}
\put(6.8,6.75){\rule{4.8pt}{0.3pt}}
\put(7.3,6.5){$\frac{1}{2\sqrt{3}}$}
\put(7.3,1.25){$\frac{-1}{\sqrt{3}}$}
\put(4.07,4.8){\rule{0.3pt}{4.8pt}}
\put(3.06,3.8){$-\frac{1}{2}$}
\put(9.93,4.8){\rule{0.3pt}{4.8pt}}
\put(9.7,3.8){$\frac{1}{2}$}
\put(28,10){$\bss{\bar{3}}$}
\put(17,5){\line(1,0){10}}
\put(22,0){\line(0,1){10}}
\put(27,4.3){$T^3$}
\put(22.2,9.6){$T^8$}
\put(22,8.5){\circle*{0.5}}
\put(20.7,8.5){m}
\put(24.93,3.25){\circle*{0.5}}
\put(25.43,2.75){c}
\put(19.07,3.25){\circle*{0.5}}
\put(19.57,2.75){y}
\put(21.8,3.25){\rule{4.8pt}{0.3pt}}
\put(22.3,3){$\frac{-1}{2\sqrt{3}}$}
\put(22.3,8.25){$\frac{1}{\sqrt{3}}$}
\put(19.07,4.8){\rule{0.3pt}{4.8pt}}
\put(18.06,5.7){$-\frac{1}{2}$}
\put(24.93,4.8){\rule{0.3pt}{4.8pt}}
\put(24.7,5.7){$\frac{1}{2}$}
\end{picture}
\end{center}
\caption[Weight diagrams of the SU(3) representations \textbf{3} and $\bss{\bar{3}}$.]{Weight diagrams of the three-dimensional representations of
  SU(3). $T^3$ and $T^8$ are the diagonal generators.}
\label{fig:su3-su3representations}
\end{figure}
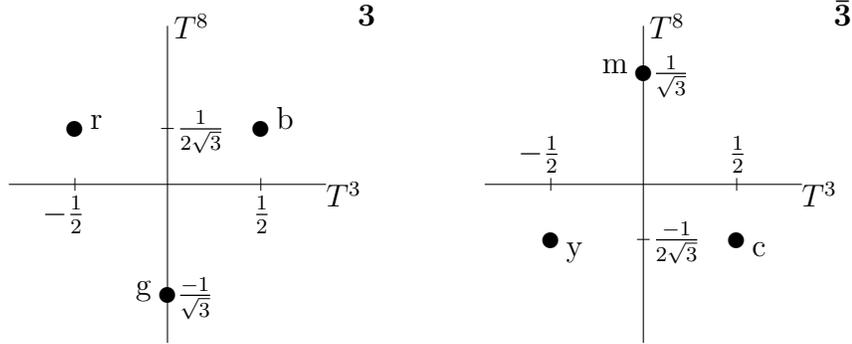

\subsection{SU(3) symmetry at $\theta=\pi/4$}
We now specify the Hilbert space the Hamiltonian~\eqref{su3}
acts on. Considering $3 \times 3 \times 3 \times 3 \times 3
...$, we can thus write the Hamiltonian more specifically in terms of
the $\lambda$s:
\begin{equation}
H=\sum_i^N \lambda_i^a \lambda_{i+1}^a.
\end{equation}
Next, note that the fundamental representation of the
Lie algebra o$(3)$ relates to the SU$(2)$ spin 1 operators, because
the $SU(2)$ is locally isomorphic to O$(3)$ and the representations
have the same dimensionality. As such, we can readily identify the
O$(3)$ type operators of the Gell Mann matrices. For $s_l$, $l=1,2,3$
and spin 1 operators, this yields 
\begin{equation}
s_1=\lambda_7, \quad s_2=-\lambda_5, \quad s_3=\lambda_2,
\end{equation}
enabling us to write the spin 1 bilinear
terms in terms of $\lambda$s. The same can be done for the biquadratic
terms, exploiting anti-commutation and commutation relations of the
Gell-Mann matrices $\lambda$~\cite{muetter}. 
%If you do it the other way around, you
%can do it like Bill Sutherland and decompose the 8 $T$'s into the $3$
%vector operator components and the $6$ traceless $2$-Tensor operators
%formed out of two spin 1 vector operators (since they are traceless,
%they totally give 8 independent operators and relate to the 8
%generators of $SU(3)$ ). In any case, you can rewrite all Lambda in
%terms of $s$ operators via some Algebra, and for the upper case you
%indeed arrive exactly at the Hamiltonian~\eqref{bb}. Actually I have
%found it easier to start from the spin 1 Hamiltonian and rewrite it
%into $\lambda$, but both of course works. A helpful reference to work
%out the calculation is K.-H. M\"utter, Zeitschrift f\"ur Physik B 96,
%105-109 (1994).
%What follows from there? As Affleck has shown, the
%Hamiltonian~\ref{su3} is gapless for general $SU(n)$. So we know that
%this spin 1 Hamiltonian out of the columnar phase must be gapless as
%well. All this is consistent with Lieb Schultz Mattis. 
%\subsection{Exact calculation}
Accordingly, the Hamiltonian part $\sum_i^N \BS_i
\cdot\BS_{i+1}$ is readily identified via
$\lambda_2=S_3, \; \lambda_5= -S_2, \; \lambda_7= S_1$ and yields
\begin{equation}
\sum_i^N \BS_i\cdot \BS_{i+1}= \sum_i^N \sum_{a=2,5,7} \lambda^a_i
\lambda^a_{i+1}. \label{su3bi}
\end{equation}
In turn, the biquadratic Hamiltonian part $H_{bq}= (\BS_i\cdot
\BS_{i+1})^2$ is rewritten as
\begin{eqnarray}
H_{bq}&=& (\lambda^2_i \lambda^2_{i+1})^2+(\lambda^5_i
\lambda^5_{i+1})^2+(\lambda^7_i \lambda^7_{i+1})^2 \label{diag}\\
&+&\lambda^2_i\lambda^5_i \lambda^2_{i+1}\lambda^5_{i+1} + [2
\leftrightarrow 5] \label{25}\\
&+&\lambda^2_i\lambda^7_i \lambda^2_{i+1}\lambda^7_{i+1} + [2
\leftrightarrow 7] \label{27}\\
&+&\lambda^5_i\lambda^7_i \lambda^5_{i+1}\lambda^7_{i+1} + [5
\leftrightarrow 7]   \label{57}
\end{eqnarray}
Whenever spatial indices are suppressed in the following, we assume that multiple powers of $\lambda$ act on the same site.
To calculate Eq.~\ref{diag}, we make use of~\ref{sym} and find 
$(\lambda^2)^2=\frac{1}{2} (\frac{4}{3} +
\frac{2}{\sqrt{3}}\lambda^8)$, $(\lambda^5)^2=\frac{1}{2} (\frac{4}{3} -
\frac{1}{\sqrt{3}}\lambda^8 +\lambda^3)$, $(\lambda^7)^2=\frac{1}{2}
(\frac{4}{3} -\frac{1}{\sqrt{3}}\lambda^8 -\lambda^3)$. Factoring
everything out, Eq.~\ref{diag} becomes $\frac{1}{2} \lambda^8_{i}
\lambda^8_{i+1}+\frac{1}{2} \lambda^3_{i} \lambda^3_{i+1}+
\frac{4}{3}$. The latter constant we can discard as a global constant
factor, the rest are just the bilinears of the Casimirs.
The subsequent lines~\ref{25},~\ref{27}, and~\ref{57}, can be calculated analogously,
so we will only show Eq.~\ref{25}. Trivial algebra yields the (anti-)symmetrized form
$\lambda^2_i\lambda^5_i \lambda^2_{i+1}\lambda^5_{i+1} + [2
\leftrightarrow 5] = \frac{1}{2}[\lambda^2,\lambda^5 ]_i
\{\lambda^2,\lambda^5 \}_{i+1}+\frac{1}{2}[\lambda^5,\lambda^2 ]_i
\{\lambda^2,\lambda^5 \}_{i+1}+\frac{1}{2}[\lambda^2,\lambda^5 ]_i
[\lambda^2,\lambda^5 ]_{i+1}+\frac{1}{2}\{\lambda^2,\lambda^5\}_i
\{\lambda^2,\lambda^5 \}_{i+1}=\frac{1}{2}[\lambda^2,\lambda^5 ]_i
[\lambda^2,\lambda^5 ]_{i+1}+\frac{1}{2}\{\lambda^2,\lambda^5\}_i
\{\lambda^2,\lambda^5 \}_{i+1}$. Now, we can make use
of~\eqref{antisymm} and~\eqref{sym} to find $[\lambda^2,\lambda^5
]=i\lambda^7$ and $\{\lambda^2,\lambda^5 \}=\lambda_6$, such that the
final result is
$\frac{1}{2}(\lambda^6_i\lambda^6_{i+1}-\lambda^7_i\lambda^7_{i+1})$. We find
\begin{equation}
\sum_i^N (\BS_i\cdot\BS_{i+1})^2=\frac{1}{2} \sum_i^N \left(\sum_{a=1,3,4,6,8}
\lambda^a_i \lambda^a_{i+1}-\sum_{a=2,5,7}
\lambda^a_i \lambda^a_{i+1}\right).\label{su3bq}
\end{equation}
Adding \eqref{su3bi} and \eqref{su3bq} yields the Hamiltonian
to be symmetric with respect to the Gell-Mann index $a$, i.e. to SU$(3)$.

\subsection{SU(3) symmetry at $\theta=3\pi/2$}
Again starting from \ref{su3}, we consider the Hilbert space spanned by $3
\times \bar{3} \times 3 \times \bar{3}....$. This has consequences for the SU$(3)$
Hamiltonian, as the imaginary-valued $\lambda$'s for $a=2,5,7$ acting on $\bar{3}$ yield one minus sign due to complex conjugation. Performing the same algebra as before, we find the purely biquadratic Hamiltonian~\eqref{su3bq} to be SU(3) invariant if acting on the modified Hilbert space,  anticipating a doubling of the unit cell along the dimerized ground state at $\theta=3\pi/2$.
%this is the Hamiltonian that relates to the canonic $SU(3)$ Heisenberg
%model. This Hamiltonian, however, has a gap! This is consistent with
%Lieb Schultz Mattis since the translational symmetry is already broken
%modulo 2 due to the Hilbert space structure.

\section{Scaling of the entanglement gap}\label{app:egap-scaling}
%%%%%%%%%%%%%%%%%%%%%%%%%%%%%%%%%%%%%%%%%
\begin{figure}[h!]
\centering
\includegraphics[width=0.95\linewidth]{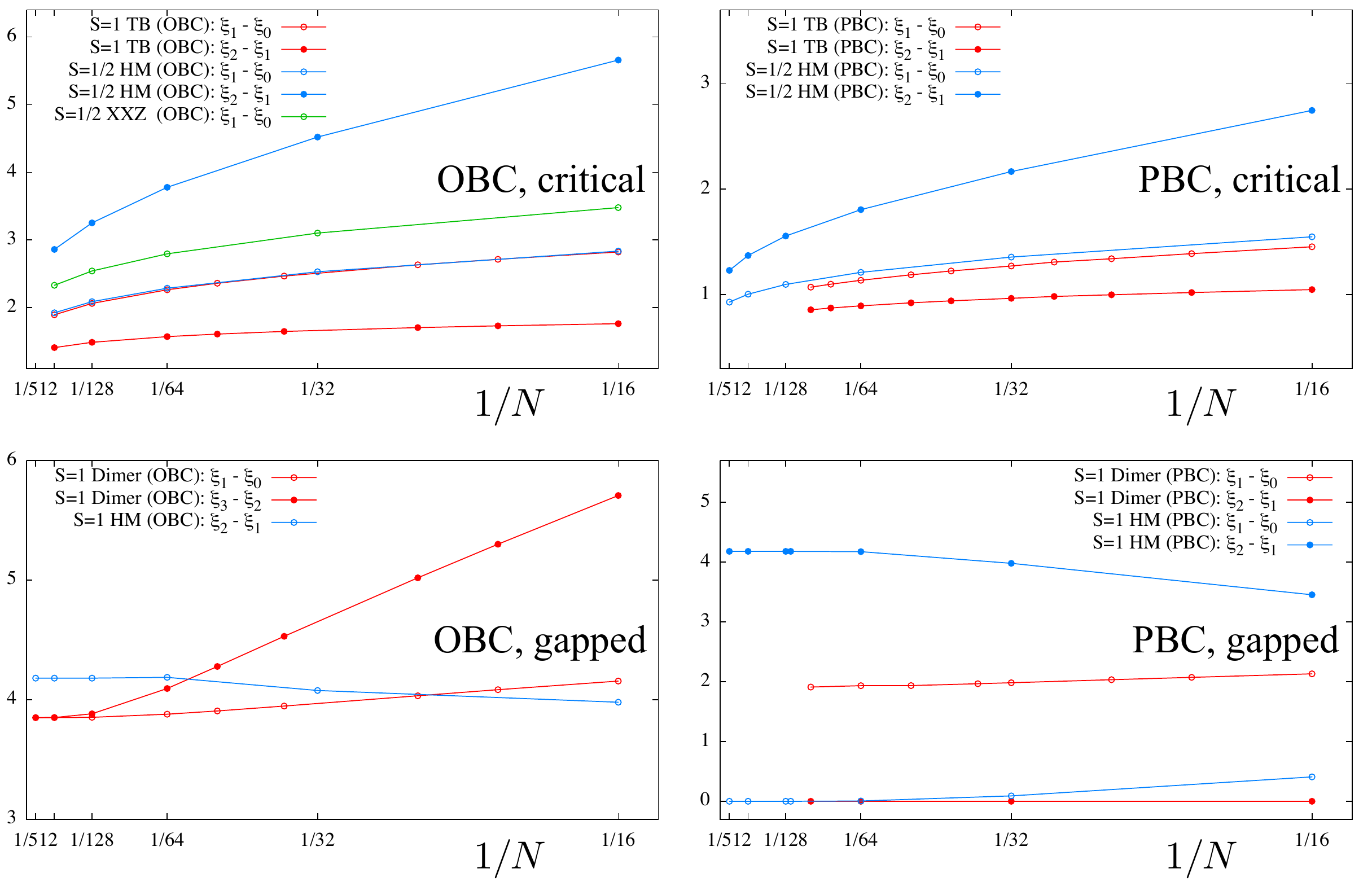}
\caption{(Color online) Scaling of the entanglement gap for gapped and gapless models for OBC and PBC obtained within DMRG. While the gapless models exhibit non-trivial spectral flow for large system sizes (see text for details), the gapped models show saturation of the entanglement gap.}
\label{fig:egap-scaling}
\end{figure}
%%%%%%%%%%%%%%%%%%%%%%%%%%%%%%%%%%%%%%%%%%

As discussed in Sec.\,\ref{sec:dimer}, the even-odd behavior of the ES with respect to entanglement cuts corresponding to even and odd subsystem lengths is reminiscent of translation symmetry breaking in the dimerized phase. Different, even though similar even-odd effects, however, can also be observed under different circumstances such as for for a gapless spin-${1\over 2}$ Heisenberg chain. For the even-odd discrepancy to provide substantiated information, it is hence important to distinguish a gapped from a gapless phase. It turns out that the scaling of the entanglement gap clearly distinguishes both cases (we restrict the discussion to half-chain cuts). We define the entanglement gap as the gap between the entanglement levels belonging to the entanglement ground state and higher-lying entanglement levels. For instance, for spin 1 Heisenberg chain with PBC the entanglement gap is above the lowest singlet and triplet; for the spin-${1\over 2}$ Heisenberg chain, we consider the gap above the single lowest singlet state. Fig.~\ref{fig:egap-scaling} shows examples for gapped and gapless models. Plots on the left (right) correspond to OBC (PBC). The spin 1 Takhtajan-Babujian (TB) model ($\theta=7\pi/4$), the spin-${1\over 2}$ Heisenberg chain, and the spin-${1\over 2}$ XXZ chain with Ising anisotropy $\Delta=0.5$ are examples for the gapless case, the spin 1 Heisenberg chain ($\theta=0$) and the dimer model with $\theta=3\pi/2$ are examples for the gapped case. For the spin-${1\over 2}$ Heisenberg model and the spin 1 TB model, both $\xi_1 - \xi_0$ and $\xi_2 - \xi_1$ are shown. The characteristic scaling behavior vs.\ $1/N$ agrees with the prediction of Calabrese and Lefevre\,\cite{calabrese-08pra032329,laflorencie-rachel}. For gapped models, we find that the entanglement gap already saturates for rather small systems sizes ($N=30$ for OBC, $N=60$ for PBC), supporting the existence of an entanglement gap in the thermodynamic limit.

% justifying to call it an entanglement {\it gap}. If we consider differences $\xi_1-\xi_0$ where both levels belong to the entanglement ground state (this typically happens for PBC) we find this difference to vanish. These findings might not be too surprising, but it clearly shows that a gapped chain also exhibits an entanglement gap (even in the absence of topological order or symmetry protected topological order) establishing another similarity between energy and entanglement spectrum.

\vspace{20pt}

%\bibliographystyle{prsty} %modernref} %plain} %prsty}
%\bibliography{s1es-JSTAT}

\end{document}